\documentclass[manuscript]{aastex}
\usepackage{aas_macros,url,natbib,amssymb,amsmath,CJK,textcomp}

\shorttitle{Dormant Comets Among the NEO Population}
\shortauthors{Ye et al.}

\begin{document}
\begin{CJK*}{UTF8}{gbsn}

\title{Dormant Comets Among the Near-Earth Object Population: A Meteor-Based Survey}

\author{Quan-Zhi Ye (叶泉志)}
\affil{Department of Physics and Astronomy, The University of Western Ontario, London, Ontario N6A 3K7, Canada}
\email{qye22@uwo.ca}

\and

\author{Peter G. Brown, Petr Pokorn\'{y}}
\affil{Department of Physics and Astronomy, The University of Western Ontario, London, Ontario N6A 3K7, Canada}

\begin{abstract}
Dormant comets in the near-Earth object (NEO) population are thought to be involved in the terrestrial accretion of water and organic materials. Identification of dormant comets is difficult as they are observationally indistinguishable from their asteroidal counterparts, however they may have produced dust during their final active stages which potentially are detectable today as weak meteor showers at the Earth. Here we present the result of a reconnaissance survey looking for dormant comets using 13~567~542 meteor orbits measured by the Canadian Meteor Orbit Radar (CMOR). We simulate the dynamical evolution of the hypothetical meteoroid streams originated from 407 near-Earth asteroids in cometary orbits (NEACOs) that resemble orbital characteristics of Jupiter-family comets (JFCs). Out of the 44 hypothetical showers that are predicted to be detectable by CMOR, we identify 5 positive detections that are statistically unlikely to be chance associations, including 3 previously known associations. This translates to a lower limit to the dormant comet fraction of $2.0\pm1.7\%$ in the NEO population and a dormancy rate of $\sim 10^{-5}~\mathrm{yr^{-1}}$ per comet. The low dormancy rate confirms disruption and dynamical removal as the dominant end state for near-Earth JFCs. We also predict the existence of a significant number of meteoroid streams whose parents have already been disrupted or dynamically removed.
\end{abstract}


\section{Introduction}

Dormant comets are comets that have depleted their volatiles and are no longer ejecting dust\footnote{We note that the term ``extinct comet'' is also frequently used in the literature. Strictly speaking, ``dormant comet'' is usually associated with comets that only temporarily lose the ability to actively sublimate, while the term ``extinct comet'' usually refers to the cometary nuclei that have permanently lost the ability to sublimate \citep[c.f.][for a more comprehensive discussion]{Weissman2002}. However, in practice, it is difficult to judge whether the comet is temporarily or permanently inactive. In this work we use the general term ``dormant comet'' which can mean either scenario.}. Due to their inactive nature, dormant comets cannot be easily distinguished from their asteroidal counterparts by current observing techniques \citep[e.g.][]{Luu1990b}. As the physical lifetime of a comet is typically shorter than its dynamical lifetime, it is logical that a large number of defunct or dormant comets exist \citep{Wiegert1999, DiSisto2009}. Dormant comets in the near-Earth object (NEO) population are of particular interest, as they can impact the Earth and contribute to the terrestrial accretion of water and organic materials as normal comets \citep[e.g.][and the references therein]{Hartogh2011f}.

It has long been known that the dust produced by Earth-approaching comets can be detected as meteor showers at the Earth \citep[e.g.][]{Schiaparelli1866, Schiaparelli1867}. Dormant comets, though no longer being currently active, may have produced dust during their final active phases, which are potentially still detectable as weak meteor showers. This has a significant implications for the investigation of dormant comets, as any cometary features of these objects are otherwise no longer telescopically observable. Past asteroid-stream searches have revealed some possible linkages, the most notable being (3200) Phaethon and the Geminids \citep[e.g.][and many others]{Williams1993e, deLeon2010, Jewitt2013} as well as (196256) 2003~EH$_1$ and the Quadrantids \citep{Jenniskens2004, Abedin2015}, both involving meteor showers that are exceptional in terms of activity. However, most showers are weak in activity, making parent identification difficult.

Radar was introduced into meteor astronomy in the 1940s and has developed into a powerful meteor observing technique \citep[c.f.][]{Ceplecha1998}. Radar detects meteors through the reflection of transmitted radio pulses from the ionized meteor trail formed during meteor ablation. Radar observations are not limited by weather and/or sunlit conditions and are able to detect very faint meteors. The Canadian Meteor Orbit Radar (CMOR), for example, has recorded about 14 million meteor orbits as of May 2016, which is currently the largest dataset for meteor orbits and hence a powerful tool to investigate weak meteor showers.

Efforts have been made to the search for dormant comets for several decades. Among the early attempts, \citet{Kresak1979e} discussed the use of the Tisserand parameter \citep{Tisserand1891} as a simple dynamical indicator for the identification of dormant comets. Assuming Jupiter as the perturbing planet, the Tisserand parameter is defined as

\begin{equation}
 T_\mathrm{J} = \frac{a_\mathrm{J}}{a} + 2 \sqrt{\frac{a(1-e^2)}{a_\mathrm{J}}} \cos{i}
\end{equation}

\noindent where $a_\mathrm{J}$ is the semi-major axis of Jupiter, and $a$, $e$, and $i$ are the semi-major axis, eccentricity, and inclination of orbital plane of the small body. A small body is considered dynamically comet-like if $T_\mathrm{J}\lesssim3$. An asteroid with $T_\mathrm{J}\lesssim3$ is classified as an asteroid in cometary orbit (ACO). Note that dormant comets and ACOs are not all physically comets originate from the Kuiper belt, as a fraction of ACOs might originate from the main asteroid belt \citep[e.g.][]{Binzel2004a}. Separation of main belt interlopers is difficult, but attempts have been made both dynamically \citep[e.g.][]{Fernandez2002ad,Tancredi2014d} and spectroscopically to separate possible cometary nuclei from asteroidal bodies \citep[e.g.][]{Fernandez2005b,DeMeo2008b,Licandro2016}. However, few attempts have been made to link ACOs with meteor showers. \citet{Jenniskens2008} provided a comprehensive review of meteoroid streams possibility associated with dormant comets based on the similarity between their orbits, but a comprehensive contemporary ``cued'' survey to look for all possible weak streams from the large number of recently discovered ACOs/NEOs that may have had weak past activity, including formation of early meteoroid trails, is yet to be performed.

In this work, we present a survey for dormant comets in the ACO component in the NEO population through the meteoroid streams they might have produced during their active phase, using the most complete CMOR dataset available to date. The survey is performed in a ``cued search'' manner rather than a commonly-used blind search: we first identify eligible ACOs (i.e. with well-determined orbits suitable for long-term integration) in the NEO population (\S~2), then simulate the formation and evolution of the meteoroid trails produced by such ACOs assuming they have recently been active (\S~3), and then search the CMOR data using the virtual shower characteristics to identify ``real'' streams now visible at the Earth (\S~4). Our survey thus simulates \textit{all} near-Earth ACOs (NEACOs) which are now known and which would have produced meteor showers at the Earth if they were recently active. This approach accounts for orbital evolution of the parent \textit{and} the subsequent evolution of the virtual meteoroid stream.

\section{Identification of Potential Shower-Producing Objects}

\subsection{Dormant Comets in the NEO Population}
\label{sec:can1}

To establish the starting conditions for the survey, we first identify possible dormant comets in the NEO population. By definition, NEOs have perihelion distance $q<1.3$~AU. In this work we focus our sampling among the orbit range of Jupiter-family comets or JFCs which overlaps the NEO population (near-Earth JFCs or NEJFCs as used by other authors). We use a slightly more relaxed constraint than the original Tisserand's derivation, namely $1.95<T_\mathrm{J}<3.05$. This is because the Tisserand parameter is derived assuming restricted three-body problem with a circular orbit of the planet; in reality, $T_\mathrm{J}$ is only an approximate, since the orbit of the planet is never strictly circular, as well as the fact that the small body is also perturbed by other planets \citep[e.g.][p.73]{Murray1999al}. We also consider the precision of the perturbed orbit solution, which is parameterized as the Uncertainty Parameter, $U$ \citep[see][]{Marsden1978}. We only consider objects with $U\leq2$, as objects falling into this category are considered ``secure'' and will be permanently numbered\footnote{See \url{http://www.minorplanetcenter.org/iau/info/UValue.html}, retrieved 2016 February 10.}. With these criteria, we identify a total of 407 objects from 13~763 known NEOs as of 2016 February 9. These 407 ACOs in the NEO population represent possible dormant comets which may have produced meteoroid streams in the recent past.

\subsection{Objects with Detectable Meteor Showers}
\label{sec:can2}

The next step is to simulate the ``virtual'' meteoroid stream of each object to see if a meteor shower is currently detectable by CMOR. Following the discussion in \citet{Ye2016c}, the meteoroid flux $\mathcal{F}$ at Earth can be calculated by

\begin{equation}
 \mathcal{F} = \frac{\eta N_\mathrm{m} \tau_\mathrm{stream}}{P^2 \Delta t_\mathrm{shower}^2 V_\oplus^2}
\label{eq:met-flux}
\end{equation}

\noindent where $\eta$ is the fraction of \textit{potentially visible meteoroids}, a subset of the Earth-bound meteoroids that may be visible as meteors, defined as meteoroids with Minimum Orbital Intersection Distance (MOID) $<0.01$~AU with respect to the Earth's orbit \citep[typical cross-section of meteoroid stream, see][]{Brown1998gi,Gockel2000a}; $N_\mathrm{m}$ is the meteoroid/dust production of the parent, which we take $N_\mathrm{m} \sim 10^{15}$ per orbit as a median case for near-Earth JFCs (elaborated in Appendix~\ref{app:median-jfc-model}); $\tau_\mathrm{stream}$ is the age of the meteoroid stream; $P$ is orbital period of the parent; $\Delta t_\mathrm{shower}$ is the duration of the meteor shower (defined by half-width-half-maximum of the shower), and $V_\oplus=30~\mathrm{km s^{-1}}$ is the orbital speed of the Earth.

\begin{figure*}
\centering
\includegraphics[width=\textwidth]{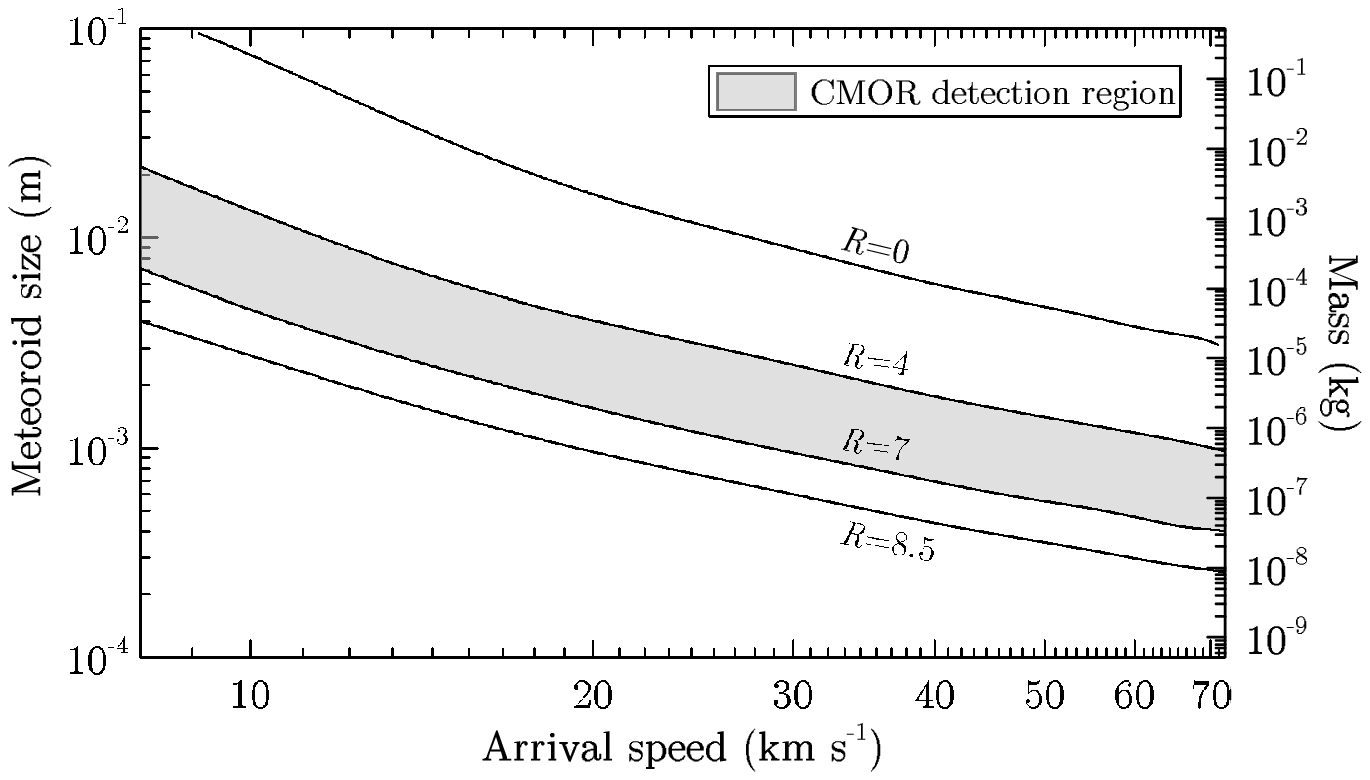}
\caption{Size-speed relation of meteors at absolute magnitude in the general $R$ bandpass of $R=0$ (typical detection limit of all-sky video networks), $R=4$ \citep[typical detection limit of narrow field video networks, as well as the upper limit of automated radar detection as meteor echo scattering changes from the underdense to the overdense regime, c.f.][]{Ye2014k}, $R=7$ (CMOR median for meteor orbits) and $R=8.5$ (CMOR detection limit) assuming bulk density of $1000~\mathrm{kg~m^{-3}}$. Calculated using the meteoroid ablation model developed by \citet{Campbell-Brown2004b}, where the luminous efficiency is constant at $0.7\%$ and the ionization coefficient is from \citet{Bronshten1981b}. Note that other authors \citep{Jones1997c,Weryk2013} have argued that these coefficients may be off by up to a factor of $\sim10$ at extreme speeds ($v_\mathrm{g}\lesssim15~\mathrm{km~s^{-1}}$ or $v_\mathrm{g}\gtrsim70~\mathrm{km~s^{-1}}$), but most of the showers we examined in this work have moderate $v_\mathrm{g}$, hence this issue does not impact our final results. The CMOR detection range is appropriated to an ionization coefficient $I$ of 5--100 in \citet{Wiegert2009}'s model.}
\label{fig:cmor-size}
\end{figure*}

\begin{figure}
\includegraphics[width=0.5\textwidth]{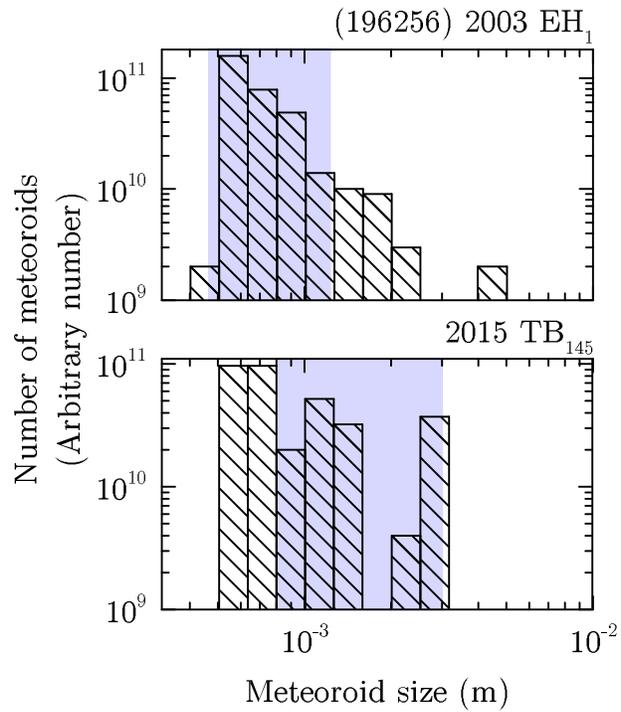}
\caption{Examples of altered arrival size distribution due to different delivery efficiency at different sizes. The meteoroids from (196256) 2003 EH$_1$ (top figure) is more similar to the original size distribution at the parent, while for the case of 2015 TB$_{145}$ (lower figure), larger meteoroids are more efficiently delivered than smaller meteoroids. Shaded areas are the CMOR-detection size range.}
\label{fig:arrival-size-example}
\end{figure}

The CMOR-observed flux will be different from $\mathcal{F}$ as the detection efficiency of CMOR is a function of the meteoroid arrival speed (Figure~\ref{fig:cmor-size}) and meteoroid stream size distribution (Figure~\ref{fig:arrival-size-example}). For each virtual stream, we assign $\eta_\mathrm{CMOR}$ being the CMOR detection efficiency, as

\begin{equation}
 \mathcal{F}_\mathrm{CMOR} = \eta_\mathrm{CMOR} \cdot \mathcal{F}
\label{eq:met-flux-cmor}
\end{equation}

There remain four unknown variables: $\eta$, $\tau_\mathrm{stream}$, $\Delta t_\mathrm{shower}$ and $\eta_\mathrm{CMOR}$. Typical numbers for the first three variables are $\eta \sim 0.1$ (i.e. 1 out of every 10 simulated meteoroids will reach the Earth), $\tau_\mathrm{stream} \sim$ a few $10^2$~yr and $\Delta t_\mathrm{shower} \sim$ a few days, but all of these quantities are highly variable \citep{McIntosh1983g,Cremonese1997d,Jenniskens2006b}. For our survey we compute all of these quantities per object (and the fourth variable, $\eta_\mathrm{CMOR}$) numerically using the following procedure.

We define $\tau_\mathrm{stream}$ first as the other three variables depend on it. Simulations are performed using the MERCURY6-based \citep[c.f.][]{Chambers1999a} meteoroid model developed in our earlier works \citep[e.g.][]{Ye2014j,Ye2015}. We employ the ejection model described by \citet{Jones1995} for meteoroids with sizes between 0.5~and 50~mm, an ``envelop'' size range appropriate to the detection range of CMOR as given by the meteoroid ablation model \citep[Figure~\ref{fig:cmor-size}, c.f.][]{Campbell-Brown2004b}, assuming a bulk density of $1~000~\mathrm{kg \cdot m^{-3}}$ and a size distribution of $\mathrm{d}N/\mathrm{d}a\propto a^{-q}$ where $q=3.6$ \citep[][\S~5]{Fulle2004}. Meteoroids are released in a time step of 10~d when the parent is within the sublimation line ($r_\mathrm{h}<2.3$~AU). The system of planets, parent bodies and meteoroids are then integrated with the RADAU integrator \citep{Everhart1985} with an initial time step of 7~d. Time step is reduced upon close encounters as documented in \citet{Chambers1997e}. Gravitational perturbations from the eight major planets (with the Earth-Moon system represented by a single mass at the barycenter of the two bodies), radiation pressure, and Poynting-Robertson effect are considered in the integration.

We operationally define $\tau_\mathrm{stream}$ as the time taken for the median $D$-parameter of any two test meteoroids to grow beyond a given threshold. The $D$-parameter was originally introduced by \citet{Southworth1963c} for meteor shower identification; it is essentially a measure of the similarity between a pair of orbits denoted as $A$ and $B$:
 
\begin{equation}
 D_{A, B}^2 = \left(q_B - q_A \right)^2 + \left( e_B - e_A \right)^2 + \left( 2\sin{ \frac{I}{2} } \right)^2 + \left[ \left(e_A + e_B \right) \sin{ \frac{\varPi}{2} } \right]^2
\end{equation}
 
\noindent where
 
\begin{equation}
 I = \arccos{ \left[ \cos{i_A} \cos{i_B} + \sin{i_A} \sin{i_B} \cos{\left( \varOmega_A - \varOmega_B \right)} \right]} 
\end{equation}
 
\begin{equation}
 \varPi = \omega_A - \omega_B + 2 \arcsin{ \left( \cos{\frac{i_A+i_B}{2}} \sin{\frac{\varOmega_A-\varOmega_B}{2}} \sec{\frac{I}{2}} \right) } 
\end{equation}
 
\noindent and the subscripts $A$ and $B$ refer to the two orbits being compared. Here $q$ is the perihelion distance in AU, $e$ is the eccentricity, $i$ is the inclination, $\Omega$ is the longitude of ascending node, and $\omega$ is the argument of perihelion. The sign of the $\arcsin$ term in the equation for $\varPi$ switches if $|\Omega_A-\Omega_B|>180^\circ$.

The physical meaning of $\tau_\mathrm{stream}$ can be interpreted as a measure of the dispersion timescale of the meteoroid stream, equivalent to the age of the stream.

For each object, the simulation starts with $\tau_\mathrm{stream}=100$~yr. This value is incremented in steps of 100~yr, until the median $D$-parameter among all test particles that composed the virtual stream reaches $D=0.1$, an empirical cutoff that was found by \citet{Southworth1963c} and was later revisited by many \citep[e.g.][]{Sekanina1976g, Drummond1981o, Ceplecha1998}; or $\tau_\mathrm{stream}=10^4$~yr which we adopt as an operational upper limit for the simulation, as this is comparable to the oldest estimated stream ages based on de-coherence timescales \citep{Pauls2005}.

Once $\tau_\mathrm{stream}$ is determined, we calculate the MOID of each test meteoroids with respect to the Earth's orbit at the epoch of 2012 Jan. 1 Terrestrial Time and collect those with MOID$<0.01$~AU (i.e. potentially visible meteoroids). The values of $\eta$, $\eta_\mathrm{CMOR}$ (the number of meteoroids detectable by CMOR size bins divided by the total number of potentially visible meteoroids) and $\Delta t_\mathrm{shower}$ (defined as the standard deviation of the solar longitudes of the MOID points) are readily available at this stage. The number of test meteoroids making the CMOR-detectable virtual meteor shower (not the total simulated test meteoroids which is $\sim10^5$) at this stage is typically $\sim10^3$ and is at least 100. The virtual meteoroid shower flux for each parent is then calculated using Eq.~\ref{eq:met-flux} and~\ref{eq:met-flux-cmor}. The detection limit for multi-year CMOR data is of the order of $10^{-3}~\mathrm{km^{-2}~hr^{-1}}$ \citep{Bruzzone2015}. Hence, we only consider virtual showers with $\mathcal{F}_\mathrm{CMOR} \gtrsim 10^{-3}~\mathrm{km^{-2}~hr^{-1}}$ as CMOR-detectable showers.

Readers may immediately notice that, for a significant fraction of the objects, the calculated $\tau_\mathrm{stream}$ is beyond the typical chaotic timescale of JFCs \citep[$\sim1000$~yrs, e.g.][]{Tancredi1995}. Here it is important to note that our approach focuses at the \textit{mean~orbit} rather than the exact position of the parent; therefore we are to examine the chaotic timescale of the orbit instead of the parent. For each object, we generate 100~clones from the covariance matrix of the orbital elements\footnote{Available from the JPL Small-Body Database, retrieved on 2016 February 15.} and integrate them backward in time. Similar to the definition of $\tau_\mathrm{stream}$, we define the parental orbital chaotic timescale $\tau_\mathrm{parent}$ as the time taken for the median $D$-parameter of any two clones to grow beyond 0.1. Thus, $\tau_\mathrm{parent}$ corresponds to the time that the parent orbit is well constrained and the associated meteoroid stream can therefore be simulated with confidence. The value of $\tau_\mathrm{stream}$ should be viewed cautiously if $\tau_\mathrm{parent} \ll \tau_\mathrm{stream}$.

Following this procedure, we identify 44 objects that meet both our visibility and detection criteria, and that the stream formation process have the potential of producing CMOR-detectable meteor activity between 2002 and 2015. Note that no geographic constraint is considered at this stage; i.e. southerly virtual radiants are still included. Detailed results are tabulated in Table~\ref{tbl:can-met-ev} and in Appendix~\ref{app:rad-act-size}. The values of $\tau_\mathrm{parent}$ and $\tau_\mathrm{stream}$ for each object are also listed. Several known asteroid-stream linkages are among the list, such as (196256) 2003~EH$_1$ -- Quadrantids \citep{Jenniskens2004,Abedin2015} and 2004 TG$_{10}$ -- Taurid complex \citep{Jenniskens2006b,Porubvcan2006}. For these established linkages, the calculated radiants and arrival speeds agree with observations within uncertainties, providing some basic validation of the meteoroid modeling approach. In particular, we note that the calculated stream ages ($\tau_\mathrm{stream}$) are 300~yr for Quadrantids and 6100~yr for the 2004 TG$_{10}$ component in the Taurid complex, consistent with previous findings \citep[200~yr for Quadrantids and $\sim10^4$~yr for Taurids, e.g.][]{Steel1991f,Abedin2015}. Additionally, our model predicts a flux of $0.012~\mathrm{km^{-2}~hr^{-1}}$ for the Quadrantids to CMOR's limiting sensitivity, broadly consistent with daily average fluxes of a few $0.01~\mathrm{km^{-2}~hr^{-1}}$ \citep[e.g.][]{Brown1998c}. The modeled flux for the Taurids, however, is about 100 times higher than observations, likely related to the formation mechanism of the Taurids not being purely sublimation-driven \citep[c.f.][\S~25, and the references therein]{Jenniskens2006b} which differs from our modeling assumption.

\section{Prediction of Virtual Meteor Showers}
\label{sec:prediction}

Meteor activity is classified into two categories: annual showers, which are visible every year at more or less the same time and rate; and outbursts, which are enhancements visible in some years but not others. This divide plainly reflects the evolution of the meteoroid cloud: recently-formed meteoroid \textit{trails} experience little differential effects due to radiation pressure and planetary perturbation, and thus tend to remain concentrated in a narrow arc in the orbit and only become visible as meteor outbursts when this ``knot'' of denser material impacts the Earth. After some time, differential effects gradually stretch the trail along the entire orbit into a meteoroid \textit{stream}, visible as an annual meteor shower every time the Earth arrives at the stream intersection point. Outbursts from young trails provide clues to the ejection state (epoch, particle ejection speed, etc.) of the trails, such as the case of 55P/Tempel-Tuttle and the Leonids \citep[e.g.][]{Yeomans1996}. In contrast, more highly evolved streams are useful for the estimation of the age of the entire stream, such as the case of 109P/Swift-Tuttle and the Perseids \citep[e.g.][]{Brown1998gi}.

For the prediction of annual showers, we use the simulation result obtained in \S~\ref{sec:can2} and calculate the radiants and timing of the potentially visible meteoroids at the Earth. Results are tabulated in Table~\ref{tbl:can-met-ev}. The values of $\tau_\mathrm{parent}$ and $\tau_\mathrm{stream}$ are also listed in the same table. We find the median $\tau_\mathrm{parent}$ to be 4300~yr, comparable to the typical timescale of $100\%$ growth in the positional uncertainty of a JFC \citep{Tancredi1995}. For individual bodies, $\tau_\mathrm{parent}$ depends on the dynamical characteristics of the body as well as the precision of the observation. Objects with extremely long $\tau_\mathrm{parent}$ are usually found in/near mean-motion resonances, and/or observed by high-precision techniques (e.g. radar observations). The median $\tau_\mathrm{stream}$ is found to be $\sim 1800$~yr, which is also consistent with other studies \citep[e.g.][\S26.1]{Babadzhanov1992h, Jenniskens2006b}.

For the prediction of meteor outbursts, we first probe the transition timescale from trail to stream, or simply the \textit{encircling~time} of the meteoroid cloud, $\tau_\mathrm{enc}$. In another sense, $\tau_\mathrm{enc}$ corresponds to the time that the ejection state of a meteoroid trail is preserved. We define $\tau_\mathrm{enc}$ as the time taken for the standard deviation of the mean anomalies of the meteoroids to reach $60^\circ$ (The mathematical consideration is that $99.7\%$ or $3\sigma$ of the meteoroids spread to half orbit or $180^\circ$ in mean anomaly assuming a Gaussian distribution). The simulation is conducted in the same manner as the simulation in \S~\ref{sec:can2}. We then follow the evolution of the meteoroid trail formed by each parent up to $\tau_\mathrm{enc}$ years preceding 2012~AD and search for encounters between the trails and the Earth in CMOR-operational years (2002--2015). For each encounter, we estimate the meteoroid flux following the method described in \citet{Ye2016} taking the median JFC model for dust production of the parent. We only consider encounters with the Earth's orbit of less than $\sim 0.002$~AU \citep[c.f. the discussion regarding the ``second space criterion'' in][]{Vaubaillon2005a} and predicted meteoroid flux $\mathcal{F}_\mathrm{CMOR}>10^{-2}~\mathrm{km^{-2}~hr^{-1}}$, the detection limit of single year CMOR data \citep{Ye2016}. Our model predicts 25 outburst events from a total of 11 objects that are potentially detectable by CMOR. These are tabulated in Table~\ref{tbl:can-met-ob}.

\section{Observational Survey of Virtual Meteor Activity}
\label{sec:obs}

The observational data for our survey is gathered by CMOR, an interferometric backscatter radar system located near London, Canada \citep[e.g.][]{Jones2005,Brown2008,Weryk2012}. CMOR consists of one main site equipped with interferometer as well as five remote receivers, all of which operate at 29.85~MHz \citep{Ye2013w}. Orbits of the meteoroids can be derived from the interferometry and the time delay for common radar echoes between various stations. Routine and continuous observation commenced in early 2002. As of early 2016, CMOR has measured $\sim13$~million meteor orbits with a corresponding representative meteor magnitude of $\sim+7$.

Meteor showers (outbursts) are defined as an enhancement in meteor rates from a certain celestial point (the \textit{radiant}) at a certain speed over a short period of time. The wavelet transform method has been demonstrated as a robust method for shower identification in radar data \citep{Galligan2000}. Here we perform the survey search using a quasi 4-dimensional Mexican hat wavelet, with the wavelet coefficient $\psi(x_0, y_0, v_{\mathrm{g}, 0})$ at celestial coordinate $(x_0, y_0)$ and speed $(v_{\mathrm{g}, 0})$ defined as:

\begin{equation}
\begin{split}
 \psi(x_0, y_0, v_{\mathrm{g}, 0}) = \frac{1}{(2\pi)^{3/2} \sigma_v^{1/2}} \int_{v_{\mathrm{g}, \mathrm{min}}}^{v_{\mathrm{g}, \mathrm{max}}} \int_{-\infty}^\infty \int_{-\infty}^\infty f\left(x, y, v_\mathrm{g}\right) \\
 \times \left[ 3 - g\left(x, y, \sigma_\mathrm{rad}\right) - h\left(v_\mathrm{g}, \sigma_v\right) \right] \\ \times \exp{\left\{-\frac{1}{2}\left[g\left(x, y, \sigma_\mathrm{rad}\right) - h\left(v_\mathrm{g}, \sigma_v\right)\right]\right\}}~\mathrm{d}x \mathrm{d}y \mathrm{d}v_\mathrm{g}
\end{split}
\end{equation}

\noindent and

\begin{equation}
g(x, y, \sigma) = \frac{(x-x_0)^2 + (y-y_0)^2}{\sigma^2}
\end{equation}

\begin{equation}
h(v_\mathrm{g}, \sigma_v) = \frac{(v_\mathrm{g} - v_{\mathrm{g},0})^2}{\sigma_v^2}
\end{equation}

\noindent where $f(x, y, v_\mathrm{g})$ is the distribution of radiants, $\sigma_\mathrm{rad}$ and $\sigma_v$ are the spatial and speed probe sizes, $x$, $y$ and $v_\mathrm{g}$ are spatial coordinates and speed in the geocentric space of observed radiants.

To enhance the signal from annual weak showers, we follow the procedure described in \citet{Bruzzone2015} and combine the entire CMOR dataset into a stacked virtual year. The data in both calendar year and stacked virtual year are divided into $1^\circ$ solar longitude bins, producing a quasi 4-dimensional data-set that is then analyzed using the wavelet technique. The wavelet transform detects only radiants within roughly one spatial/speed probe size, as these contribute significantly to the wavelet coefficient. As such, radiant distributions that match the specified spatial/speed probe sizes will show enhanced wavelet coefficient. For most showers, the simulated radiants are very compact such that the spatial/speed spreads are comparable to or smaller than the CMOR's measurement uncertainty. For these cases we use the the empirical probe sizes of $4^\circ$ and $10\%$ adopted by \citet{Brown2008} for shower detection.

For each shower/outburst, we inspect the variation of $\psi(x_0, y_0, v_{\mathrm{g}, 0})$ as a function of time within the virtual/natural year to search for enhancements. Positive detections behave as a rise in $\psi(x_0, y_0, v_{\mathrm{g}, 0})$ that is well above the background noise \citep[e.g.][Fig. 1]{Brown2010}.

\section{Results and Discussion}

\subsection{Annual Showers from Old Streams}

\begin{figure*}
\includegraphics[width=\textwidth]{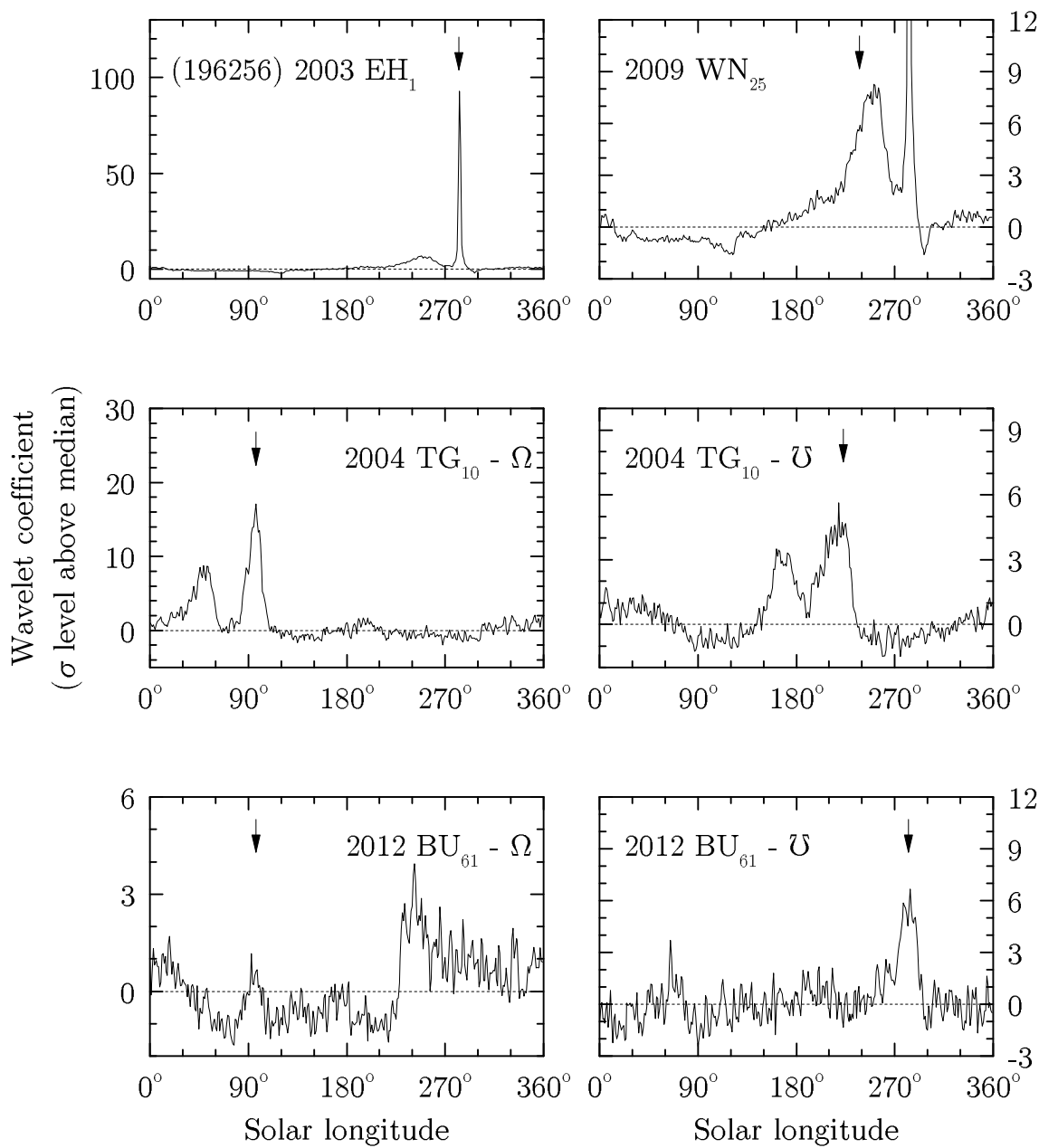}
\caption{Detection of annual meteor activity that may be associated with (196256) 2003 EH$_1$, 2004 TG$_{10}$ (both ascending node \textohm~and descending node \textmho), 2009 WN$_{25}$, 2011 BE$_{38}$ and 2012 BU$_{61}$ (both ascending node \textohm~and descending node \textmho). Activity peaks are highlighted by arrows. The figures show the relative wavelet coefficients at radiants given in each graph in units of the numbers of standard deviations above the annual median.}
\label{fig:cmor-stack-obs}
\end{figure*}

Among the 44 virtual streams predicted to be detectable by CMOR, we identify four probable positive detections in the stacked CMOR data that can be associated with (196256) 2003 EH$_1$, 2004 TG$_{10}$, 2009 WN$_{25}$, and 2012 BU$_{61}$, as shown in Figure~\ref{fig:cmor-stack-obs}. Among these associations, two are considered established: (196256) 2003 EH$_1$ to the Quadrantids \citep[e.g.][]{Jenniskens2004, Abedin2015} and 2004 TG$_{10}$ as part of the Taurid complex \citep[e.g.][]{Jenniskens2006b, Porubvcan2006}; one is recently proposed: 2009 WN$_{25}$ to the November i Draconids \citep{Micheli2016a}. For all these three cases, the predicted shower characteristics are consistent with the observations, except for the activity duration of the November i Draconids -- 2009 WN$_{25}$ pair. The predicted duration is about 1 day while the observed activity lasted for $\sim20$~days \citep{Brown2010}. This simply reflects that we assume the operational stream age $\tau_\mathrm{stream}=100$~yr while the actual stream might be much older (and thus more dispersed). One node of the detection associated with 2012 BU$_{61}$ is identified with the Daytime $\xi$ Sagittariids (descending node), which in turn has been previously associated with 2002 AU$_5$ \citep{Brown2010} though the linkage is not considered well established. The ascending nodal intersection for 2012 BU$_{61}$ can not be identified with any known showers, but does show detectable enhancement as shown in Figure~\ref{fig:cmor-stack-obs}.

A complicating issue in the parent-shower linkage is the likelihood of chance alignment. This is especially true as there are over 14~000 known NEOs and $\sim 700$ identified/proposed meteor showers as of May 2016\footnote{\url{http://www.minorplanetcenter.net/mpc/summary} and \url{http://www.astro.amu.edu.pl/~jopek/MDC2007/Roje/roje_lista.php?corobic_roje=0&sort_roje=0}, retrieved 2016 May 7.}. Therefore, \textbf{it is not sufficient to propose a linkage by simply noting the similarity of the respective orbits}. Instead, following the exploration by \citet{Wiegert2004e}, we evaluate the following question to establish orbital similarity significance: consider the $D_\mathrm{SH}$ parameter between the proposed parent-shower pair to be $D'_0$, what is the expected number of parent bodies $\langle X \rangle$ that have orbits such that $D'<D'_0$ (where $D'$ is the $D_\mathrm{SH}$ parameter between the ``new'' parent and the shower)?

This question can be answered using a NEO population model providing the orbits of the possible parent and the meteoroid stream are well known. We employ the de-biased NEO model developed by \citet{Greenstreet2012a} and generate two synthetic NEO populations down to absolute magnitude $H=18$ and $H=22$ following $\alpha=0.35$ for $H<18$ and $\alpha=0.26$ for $18<H<22$ \citep[where $\alpha$ is the size distribution index of the NEO population; see][]{Jedicke2015}. Orbits of the meteoroid streams of interest are calculated from the respective wavelet maxima as found in the CMOR data. The ascending nodal activity for 2012 BU$_{61}$ is heavily contaminated by sporadic activity later in the year ($\lambda_\odot \sim 240^\circ$) which prevents useful orbits to be obtained. This procedure is repeated for the proposed linkages of November i Draconids -- 2009 WN$_{25}$ and Daytime $\xi$ Sagittariid -- 2002 AU$_5$ and 2012 BU$_{61}$. The results are summarized in Table~\ref{tbl:met-ev-positive}.

We observe the following:

\begin{enumerate}
 \item The statistical model supports 2009 WN$_{25}$ as the likely parent for the November i Draconids.
 \item The case of 2012 BU$_{61}$ and the Daytime $\xi$ Sagittariids is complicated. The orbits derived from this work and \citet{Brown2010} is notably different from the one initially proposed in the Harvard Radio Meteor Project \citep[][listed as $\xi$ Sagittariids, though the IAU catalog has identified it as the same shower]{Sekanina1976g} which has $D_\mathrm{SH}=0.28$, though this work and \citet{Brown2010} use virtually the same data. The Daytime $\xi$ Sagittariids has not been reported by a third observing system. However, we note that the Daytime Scutids, another unestablished shower reported by the Harvard survey, resembles the orbit of the Daytime $\xi$ Sagittariids observed by CMOR \citep[see][the orbit of Daytime Scutids is appended in Table~\ref{tbl:met-ev-positive}]{Sekanina1973b}, with $D_\mathrm{SH}=0.15$. We suspect that two different showers have been accidentally assigned the same name. The association to 2012 BU$_{61}$ would be statistically significant, either using the CMOR orbit or the Harvard orbit for the Daytime Scutids.
 \item The linkage between 2002 AU$_5$ and Daytime $\xi$ Sagittariids or Daytime Scutids is not statistically significant.
\end{enumerate}

We note that among the four parent-shower associations found by our survey, three parents, namely 2004 TG$_{10}$, 2009 WN$_{25}$ and 2012 BU$_{61}$, are sub-kilometer bodies. Since sub-kilometer comets are effectively eliminated by rotational disruption \citep{Rubincam2000d, Taylor2007fj, Jewitt2010k}, we think that these four bodies are likely larger fragments from previous break-ups. In fact, 2004 TG$_{10}$ is generally recognized as being part of the Taurid complex \citep{Porubvcan2006}, while the November i Draconid streams (which 2009 WN$_{25}$ is linked to) has been considered to be associated with the Quadrantid stream \citep{Brown2010}.

In addition to the positive detections, we have not reproduced a number of previously proposed associations. Our initial shortlist included most of the objects in earlier proposed associations except objects with short orbital arc (i.e. low orbit quality). The calculation of $\langle X \rangle$ is repeated for every proposed association. As shown in Table~\ref{tbl:negative}, only 8 out of 32 previously proposed associations have $\langle X \rangle \ll 1$:

\begin{enumerate}
 \item Corvids -- (374038) 2004 HW. Linkage first proposed by \citet{Jenniskens2006b}. The Corvid meteor shower is one of the slowest known meteor showers, with $v_\mathrm{g}=9~\mathrm{km~s^{-1}}$. It was only observed in 1937 \citep{Hoffmeister1948} until being recently recovered by \citet{Jenniskens2016} and has not been detected by many radar and photographic surveys. The Corvids are undetected by CMOR, which is unsurprising as back-scatter radars are insensitive to very slow meteors.
 \item $\psi$ Cassiopeiids -- (5496) 1973 NA. Linkage first proposed by \citet{Porubcan1992e}. The object is not included in Table~\ref{tbl:can-met-ev} due to low expected flux being below the CMOR detection limit. Our test simulation shows that only a small fraction ($<0.1\%$) of sub-millimeter-sized meteoroids ($\sim 0.1$~mm) released in the past 1000~yr would be arriving at the Earth's orbit. The fact that the meteor shower is detectable by video techniques (which only detect larger, millimeter-sized meteoroids) is incompatible with the modeling result.
 \item 66 Draconids -- 2001 XQ. The unconfirmed shower has only been reported by \citet{vSegon2014d} who also propose the linkage. Our survey wavelet analysis at the reported radiant of the 66 Draconids did not detect any enhancement.
 \item $\delta$ Mensids -- (248590) 2006 CS. Linkage first proposed by \citet{Jenniskens2006b}. The unconfirmed shower is only accessible by observers in the southern hemisphere.
 \item $\iota$ Cygnids -- 2001 SS$_{287}$. Linkage first proposed by \citet{Andreic2013a} who remains the only observer of this unconfirmed shower at the time of writing. No enhancement is seen in the CMOR wavelet analysis at the reported radiant.
 \item $\kappa$ Cepheids -- 2009 SG$_{18}$. Shower discovered by \citet{vSegon2015} who also propose the linkage. The predicted radiant is consistent with the reported radiant of $\kappa$ Cepheids. The meteoroid speed is favorable for radar detection ($v_\mathrm{g}=34~\mathrm{km~s^{-1}}$) and a relatively strong flux is predicted ($\mathcal{F}=0.22~\mathrm{km^{-2}~hr^{-1}}$), however no enhancement is seen in the wavelet analysis at either the predicted or the reported radiant.
 \item Northern $\gamma$ Virginids -- 2002 FC. Linkage proposed by \citet{Jenniskens2006b}. This unestablished shower has only been reported by \citet{Terentjeva1989a} who analyzed photographic fireball observations from 1963 to 1984. No enhancement is seen in wavelet analysis of the CMOR data at the reported radiant.
 \item $\zeta^1$ Cancrids -- 2012 TO$_{139}$. Shower detection as well as potential linkage are both identified by \citet{vSegon2014g}. No enhancement is seen in the CMOR wavelet analysis at the predicted radiant. Also, the model predicts a stronger descending nodal shower which is also not seen in CMOR data.
\end{enumerate}

It should be emphasized that the statistical test only addresses the likelihood of finding a better parent body match for a given stream orbit; it does not take into account the false positives in shower identification, a complicated issue heavily investigated for half a century \citep[e.g.][and many others]{Southworth1963c,Drummond1981o,Galligan2001b,Brown2008,Moorhead2016}. There exists a danger of assigning a small body as the ``parent'' of some random fluctuation in the meteoroid background. This is especially true for unestablished showers, as most of them have been observed by only one observer.

\subsection{Outbursts from Young Trails}

\begin{figure*}
\includegraphics[width=\textwidth]{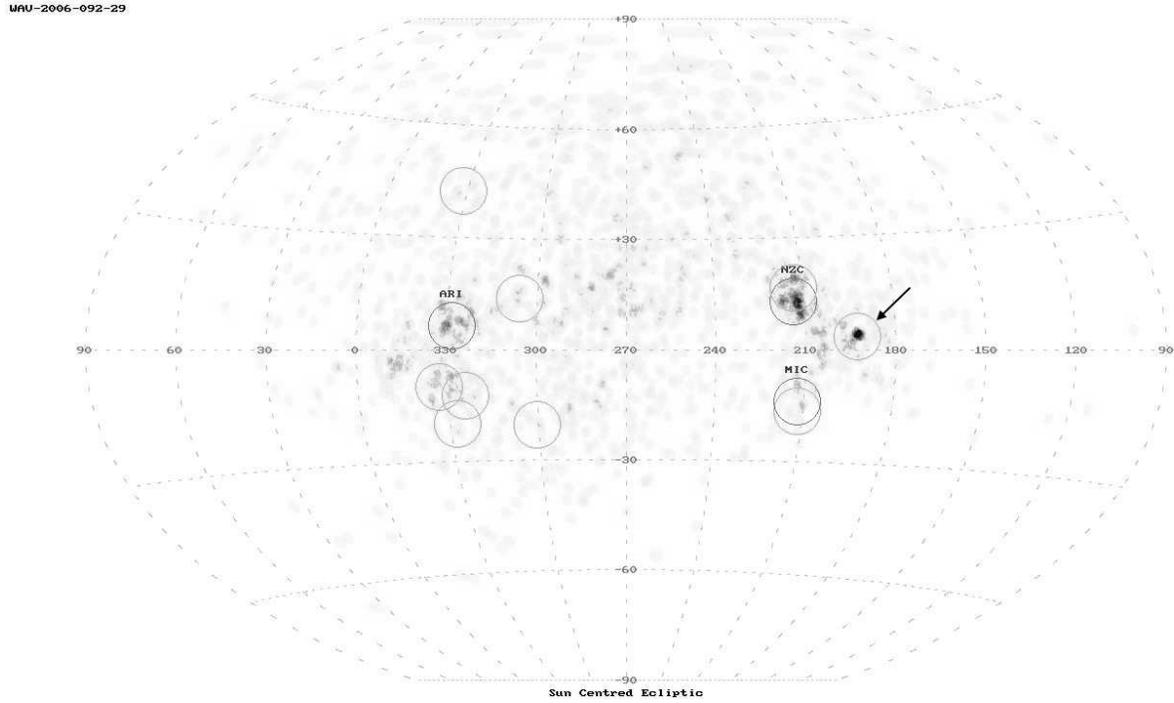}
\caption{Possible activity from (139359) 2001 ME$_1$ on 2006 Jun. 24 in sun-centered ecliptic sphere. Darker contour corresponds to areas in the sky with denser radiants. Known showers are marked by dark circles and the International Astronomical Union (IAU) shower designation (ARI = Arietids, NZC = Northern June Aquilids, MIC = Microscopiids). Unknown enhancements are marked by gray circles. Note that most enhancements are random fluctuations. The possible activity associated with (139359) 2001 ME$_1$ is the strong enhancement near $\lambda-\lambda_\odot=190^\circ$, $\beta=+5^\circ$.}
\label{fig:wav-139359}
\end{figure*}

\begin{figure*}
\includegraphics[width=\textwidth]{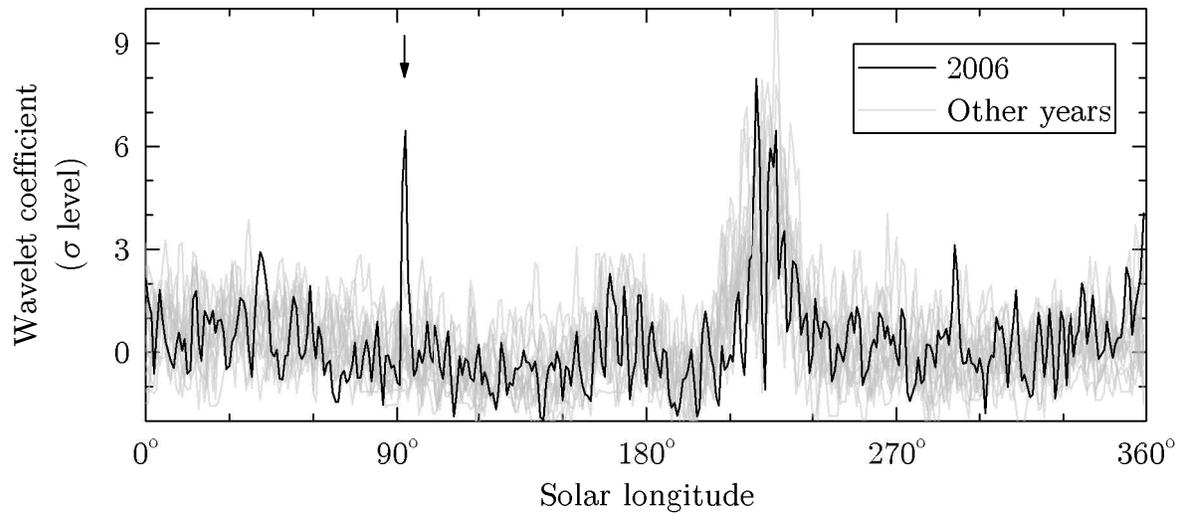}
\caption{Variation of the wavelet coefficient at $\lambda-\lambda_\odot=191^\circ$, $\beta=+4^\circ$ and $v_\mathrm{g}=30.0~\mathrm{km~s^{-1}}$ in 2002--2015 (gray lines except for 2006). Possible activity from (139359) 2001 ME$_1$ in 2006 is marked by an arrow. Recurring activity around $\lambda_\odot=220^\circ$ is from the Taurids complex in November.}
\label{fig:cmor-139359-2006}
\end{figure*}

Among the predictions given in Table~\ref{tbl:can-met-ob}, only one prediction is associated with a distinct detection: the event from (139359) 2001 ME$_1$ in 2006 (Figure~\ref{fig:wav-139359},~\ref{fig:cmor-139359-2006} and Table~\ref{tbl:cmor-139359-2006}). The association is of high statistical significance, as $\langle X \rangle_{H<18}\sim0.01$. From the wavelet profile, we estimate that the observed flux is close to the detection threshold or $\sim 10^{-2}~\mathrm{km^{-2}~hr^{-1}}$, as the signal is not very significantly higher than the background fluctuation. The event, if indeed associated with (139359) 2001 ME$_1$, should have originated from a relatively recent ($<100$~yr) ejection event. Since the observed flux is about the same order as the model prediction, it can be estimated that the dust production associated to the ejection is comparable to the average dust production of known near-Earth JFCs. Curiously, the annual shower associated with (139359) 2001 ME$_1$, though with a moderate expected flux, is not detected. This may suggest that the ejection was a transient event rather than a prolonged one, possibly similar to the activity of 107P/(4015) Wilson-Harrington upon its discovery in 1949 \citep[c.f.][]{Fernandez1997ab}.

Another interesting aspect of our survey is the negative detection of several strong predicted events (with $\mathcal{F}_\mathrm{CMOR}\gtrsim1~\mathrm{km^{-2}~hr^{-1}}$). These can be used to place a tight constraint on the past dust production of the parent. It can be concluded that the dust production of 2001 HA$_4$, 2012 TO$_{139}$ and 2015 TB$_{145}$ are either at least 2 magnitudes lower than the median near-Earth JFC model or have a much steeper dust size distribution than we assume.

\subsection{Discussion}

With the results discussed above, we now revisit the population statistics of the dormant comets. We first consider the number of streams detectable by CMOR, $\mathcal{N}_\mathrm{CMOR}$, to be expressed as 

\begin{equation}
 \mathcal{N}_\mathrm{CMOR} = \mathcal{N}_\mathrm{dc} \cdot \eta_\mathrm{NEACO} \cdot \eta_\mathrm{shr} \cdot \eta_\mathrm{CMOR}
\end{equation}

\noindent where $\mathcal{N}_\mathrm{dc}$ is the true (de-biased) number of dormant comets in the NEACO population, $\eta_\mathrm{NEACO}$ is the detection efficiency of the NEACOs (i.e. the number of known NEACOs divided by the number of total NEACOs predicted by NEO population model), $\eta_\mathrm{shr}$ is the selection efficiency of NEACOs that produce visible meteor showers (i.e. the number of shower-producing NEACOs divided by the total number of NEACOs), and $\eta_\mathrm{CMOR}$ is the detection efficiency of CMOR (i.e. the number of total virtual showers observable by CMOR divided by the total number of virtual showers visible at the Earth). Rearranging the terms, we have

\begin{equation}
 \mathcal{N}_\mathrm{dc} = \mathcal{N}_\mathrm{CMOR} \cdot \left( \eta_\mathrm{NEACO} \cdot \eta_\mathrm{shr} \cdot \eta_\mathrm{CMOR} \right)^{-1}
\end{equation}

We focus on annual shower detection in the following as the statistics for outburst detection (only 1) is too low. As presented above, we have $\mathcal{N}_\mathrm{CMOR}=2$ for $H<18$ population and $\mathcal{N}_\mathrm{CMOR}=4$ for $H<22$ population. For the remaining three coefficients:

\begin{enumerate}
 \item For $\eta_\mathrm{NEACO}$, we obtain the true (de-biased) number of NEACOs by using \citet{Greenstreet2012a}'s NEO population model and incorporate the population statistics from \citet{Stuart2001f}, \citet{Mainzer2011x} and \citet{Jedicke2015}. We derive $\mathcal{N}_\mathrm{NEACO}=200\pm30$ for $H<18$ and $2100\pm300$ for $H<22$. Considering that we have selected 407 NEACOs in our initial sample, of which 199 are bodies with $H<18$, we obtain $\eta_\mathrm{NEACO}=1^{+0.00}_{-0.13}$ for $H<18$ and $\eta_\mathrm{NEACO}=0.19_{-0.02}^{+0.03}$ for $H<22$.
 \item For $\eta_\mathrm{shr}$, we have 44 hypothetical showers as listed in Table~\ref{tbl:can-met-ev}, among which 15 are from $H<18$ bodies. This yields $\eta_\mathrm{shr} \sim 0.1$.
 \item For $\eta_\mathrm{CMOR}$, we exclude the meteoroid streams whose are either too slow for reliable radar detection \citep[$v_\mathrm{g}<15~\mathrm{km~s^{-1}}$][]{Weryk2013} or have radiants too far south for CMOR to detect ($\beta<-30^\circ$). This leaves 36 streams in Table~\ref{tbl:can-met-ev}, including 14 originating from $H<18$ bodies. This translates to $\eta_\mathrm{CMOR} \sim 0.8$.
\end{enumerate}

With all these numbers, we obtain $\mathcal{N}_\mathrm{dc}=25\pm21$ for $H<18$ population and $\mathcal{N}_\mathrm{dc}=263\pm173$ for $H<22$ population, with uncertainties derived by error propagation. This translates to a fraction of $\sim 10\%$ of dormant comets in the NEACO population independent of size. Assuming dormant comets in asteroidal orbits (i.e. $T_\mathrm{J}>3$ bodies) are negligible, we further derive a dormant comet fraction of $2.0\pm1.7\%$ for the entire NEO population, which should be considered as a lower limit. This number is at the low end of previous estimates by \citet{Bottke2002c}, \citet{Fernandez2005b} and \citet{Mommert2015b} who give ranges of 2--$14\%$. It should be noted that all these authors also assume that dormant comets in asteroidal orbits are negligible during the derivation of the dormant comet fraction.

There are two caveats in our work that may lead to an underestimation of the dormant comet fraction. Since we used the median of the dust production of \textit{known} JFCs to feed the meteoroid flux model, if the actual JFC dust production is in fact lower, our treatment will lead to an overestimation of the number of visible showers, which will in turn reduce the derived dormant comet fraction. For a dormant comet fraction of $\sim 8\%$, we need to reduce $\eta_\mathrm{shr}$ by a factor of $8\%/2\%=4$, equivalent to using a dust production model that is 10 times lower than the current median model. This is qualitatively consistent with the recent trend that more weakly-active comets are being discovered as more sensitive NEO surveys become operational. Another caveat is that the actual dust size distribution $q$ may be different than what is used in the dust model. A steeper size distribution will result in a proportionally larger number of smaller meteoroids, making the stream more dispersed and hence more difficult to be detected. This hypothesis is not supported by reported cometary observations which are found to have $q\sim3.6$ at $\micron$-range sizes \citep[e.g.][]{Fulle2004}, but a discrepancy between millimeter to sub-millimeter-sized meteoroids is possible, such as the case of 21P/Giacobini-Zinner and the Draconids \citep{Ye2014k}.

Another fundamental question is, are dormant comets in asteroidal orbits really negligible? There is at least one prominent counter-example: (3200) Phaethon ($T_\mathrm{J}=4.508$). Phaethon is associated with the Geminid meteor shower and still possesses some outgassing activity at perihelion \citep[e.g.][]{Jewitt2010j}. Nevertheless, the fact that we do not see a lot of active NEAs suggests that such objects may not be very common.

Taking the median dynamical lifetime of near-Earth JFCs to be a few $10^3$~yr, the derived dormancy rate translates to a dormancy probability of $\sim 10^{-5}~\mathrm{yr^{-1}}$ per comet independent of sizes. This is consistent with previous model predictions, and about 5--40 times lower than the disruption probability \citep{Fernandez2002ad, Belton2014a}. This result echoes earlier suggestions that near-Earth JFCs are more likely to be disrupted rather than achieving dormancy \citep{Belton2014a}. Since the typical timescale for JFC disruption, a JFC's dynamical lifetime in the NEO region, and dispersion lifetime for a resulting meteoroid stream are all at the same order (a few $10^3$~yr), there should exist a significant number of meteoroid streams with parents that are either disrupted or have been dynamically removed such that no parent can be found, supporting the speculation of \citet{Jenniskens2010k} that many meteoroid streams are produced from disrupted comets. Since disruption and dynamical removal are competing mechanisms to eliminate JFCs from the NEO region, it may be difficult to investigate the formation of these ``orphan'' streams in the absence of an observable parent.

Finally, we compare our list against the dormant comet candidates proposed in previous works. The largest list of dormant candidates to-date is published by \citet{Tancredi2014d} and includes 203 objects that resemble JFC orbits and match a set of restrictive criteria. According to our simulations, only 3 of these 203 objects have the potential of producing CMOR-detectable activity [(196256) 2003 EH$_1$, 1999 LT$_1$, and 2004 BZ$_{74}$; note that not all of the objects are in our initial shortlist, as Tancredi's list has a more relaxed constraint on orbital precision], and only 1 out of the 7 objects has a detectable shower [(196256) 2003 EH$_1$]. \citet{Kim2014a} compiled a list of 123 NEACOs and have thermal observations, 29 of which overlap with Tancredi's list. Among these, 3 have the potential of producing CMOR-detectable activity [(307005) 2001 XP$_1$, 2001 HA$_4$, and 2010 JL$_{33}$] but none of them has detectable shower. \citet{DeMeo2008b} analyzed spectra of 49 NEOs that resemble cometary orbits (excluding 6 objects that have been later identified as comets), 6 of which may produce CMOR-detectable activity [(3360) Syrinx, (16960) 1998 QS$_{52}$, (137427) 1999 TF$_{211}$, (139359) 2001 ME$_1$, (401857) 2000 PG$_3$, and 1999 LT$_1$], only one of them produces a detectable shower, (139359) 2001 ME$_1$, which they reported an albedo of 0.04 and classified as a P-type asteroid. Other works \citep[see][and the references therein]{Mommert2015b,Licandro2016} have reported smaller samples consisting of the same objects already discussed that may produce showers from our simulations. Each author has selected a somewhat different set of candidates, but there are several objects that are selected by more than one author, including (248590) 2006 CS (which is the possible parent of $\delta$ Mensid meteor shower), (394130) 2006 HY$_{51}$, (436329) 2010 GX$_{62}$, (451124) 2009 KC$_3$, 1999 LT$_1$, 2001 HA$_4$ and 2010 JL$_{33}$, none of which [except for (248590) 2006 CS, (436329) 2010 GX$_{62}$, (451124) 2009 KC$_3$ which produce southerly radiants that are difficult to detect for CMOR] has detectable meteor activity. On the other hand, the remaining three objects with detected showers (2004 TG$_{10}$, 2009 WN$_{25}$ and 2012 BU$_{61}$) are not selected by any of the surveys largely due to the lack of astrometric/physical observations, though the most recent NEOWISE catalog release includes the measurement for 2004 TG$_{10}$ \citep{Nugent2015k}.

\section{Conclusion}

We conducted a direct survey for dormant comets in the ACO component in the NEO population. This was done by looking for meteor activity originated from each of the 407 NEOs as predicted by meteoroid stream models. This sample represents $\sim80\%$ and $\sim 46\%$ of known NEOs in JFC-like orbits in the $H<18$ and $H<22$ population respectively.

To look for the virtual meteoroid streams predicted by the model, we analyzed 13~567~542 meteoroid orbits measured by the Canadian Meteor Orbit Radar (CMOR) in the interval of 2002--2016 using the wavelet technique developed by \citet{Galligan2000} and \citet{Brown2008} and test the statistical significance of any detected association using a Monte Carlo subroutine. Among the 407 starting parent bodies, we found 36 virtual showers that are detectable by CMOR. Of these, we identify 5 positive detections that are statistically unlikely to be chance association. These include 3 previously known asteroid-stream associations [(196256) 2003 EH$_1$ -- Quadrantids, 2004 TG$_{10}$ -- Taurids, and 2009 WN$_{25}$ -- November i Draconids], 1 new association (2012 BU$_{61}$ -- Daytime $\xi$ Sagittariids) and 1 new outburst detection [(139359) 2001 ME$_1$]. Except for the case of (139359) 2001 ME$_1$, which displayed only a single outburst in 2006, all other shower detections are in form of annual activity. 

We also examined 32 previously proposed asteroid-shower associations. These associations were first checked with a Monte Carlo subroutine, from which we find only 8 associations are statistically significant. Excluding 3 associations that involve observational circumstances unfavorable for CMOR detection (e.g. southerly radiant or low arrival speed), 4 out of the remaining 5 associations involve showers that have only been reported by one study, while the last association [$\psi$ Cassiopeiids -- (5496) 1973 NA] involves some observation--model discrepancy. We leave these questions for future studies.

Based on the results above, we derive a lower limit to the dormant comet fraction of $2.0\pm1.7\%$ among all NEOs, slightly lower than previous numbers derived based on dynamical and physical considerations of the parent. This number must be taken with caution as we assume a median dust production from \textit{known} JFC comets. The typical dust production of already-dead comets, however, is not truly known. A dormant comet fraction of $\sim 8\%$ as concluded by other studies would require a characteristic dust production about $10\%$ of the median model. Another caveat is the possibility of overestimating the number of visible showers (hence, reducing the derived dormant comet fraction) due to very steep dust size distribution ($q\gg3.6$), but this is not supported by cometary observations.

We also derive a dormancy rate of $\sim 10^{-5}~\mathrm{yr^{-1}}$ per comet, consistent with previous model predictions and significantly lower than the observed and predicted disruption probability. This confirms disruption and dynamical removal as the dominant end state for near-Earth JFCs, while dormancy is relatively uncommon. We predict the existence of a significant number of ``orphan'' meteoroid streams where parents have been disrupted or dynamically removed. While it is challenging to investigate the formation of these streams in the absence of an observable parent, it might be possible to retrieve some knowledge of the parent based on meteor data alone.

\acknowledgments

We thank an anonymous referee for his/her careful review. We also thank Paul Wiegert for comments and permission to use his computational resource, as well as Zbigniew Krzeminski, Jason Gill, Robert Weryk and Daniel Wong for helping with CMOR operations. This work was made possible by the facilities of the Shared Hierarchical Academic Research Computing Network (SHARCNET:www.sharcnet.ca) and Compute/Calcul Canada. Funding support from the NASA Meteoroid Environment Office (cooperative agreement NNX15AC94A) for CMOR operations is gratefully acknowledged.

\appendix

\section{Median Dust Production Model for Typical Jupiter-family Comets}
\label{app:median-jfc-model}

To estimate the dust production of a typical Jupiter-family Comet (JFC), we compile the $Af\rho$ of 35 near-Earth JFCs from the Cometas-Obs database, using the measurements provided by various observers in the duration of 2009--2015. $Af\rho$ is an indicator of the dust product of a comet \citep[c.f.][]{Ahearn1984,AHearn1995h} and is defined by

\begin{equation}
Af\rho = \frac{4r_\mathrm{h}^2\varDelta^2}{\rho} \frac{\mathcal{F}_{C}}{\mathcal{F}_{\odot}}
\end{equation}

\noindent where $r_\mathrm{h}$ is the heliocentric distance of the comet in AU, $\varDelta$ is the geocentric distance of the comet (in the same unit of $\rho$, typically in km or cm), and $\mathcal{F}_{C}$ and $\mathcal{F}_{\odot}$ are the fluxes of the comet within the field of view as observed by the observer and the Sun at a distance of 1~AU. The photometric aperture size, or $2\rho/\varDelta$, is usually determined by the threshold value that the flux reaches an asymptote.

Since $Af\rho$ measurements are conducted at various $r_\mathrm{h}$, we scale each $Af\rho$ measurement to $r_\mathrm{h}=1$~AU by

\begin{equation}
Af\rho_0 = Af\rho \cdot r_\mathrm{h}^{4/3}
\end{equation}

\begin{figure}
\includegraphics[width=0.5\textwidth]{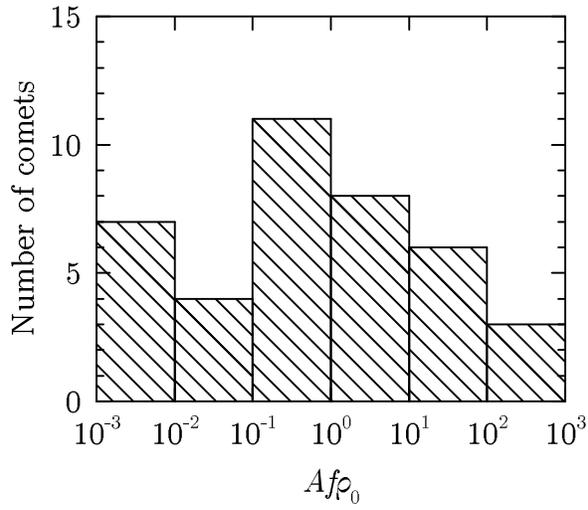}
\caption{Distribution of $Af\rho_0$ of a number of near-Earth JFCs. The median $Af\rho_0$ is 0.2~m, corresponding to a dust production rate of $7\times10^{14}$ meteoroids (appropriated to the size range of 0.5--50~mm) per orbit.}
\label{fig:jfc-afrho}
\end{figure}

The $Af\rho_0$ number can be converted to the dust production rate at 1~AU following the derivation of \citet{Ye2014p}:

\begin{equation}
N_\mathrm{d}(a_1,a_2)=\frac{655 A_1(a_1,a_2) Af\rho_0}{8 \pi A_\mathrm{B} \phi(\alpha)[A_3(a_1,a_2)+1000A_{3.5}(a_1,a_2)]}
\end{equation}

\noindent where $[a_1, a_2]$ is the size range of the meteoroids responsible for the detected flux where we use $a_1=10^{-5}$~m, $a_2=10^{-2}$~m, $A_x=(a_2^{x-s}-a_1^{x-s})/(x-s)$ for $x\neq s$ and $A_x=\ln{a_2/a_1}$ for $x=s$, with $s=3.6$ being the size population index of the meteoroids \citep{Fulle2004}, $A_\mathrm{B}=0.05$ is the Bond albedo and $\phi(\alpha)$ is the normalized phase function, which $\phi(\alpha)=1$ for isotropic scattering. From Figure~\ref{fig:jfc-afrho}, we derive median $Af\rho_0$ to be 0.2~m, corresponding to a dust production rate of $7\times10^{14}$ particles per orbit assuming an active time per orbit of $\sim1$~yr.

\section{Radiants, Activity Profiles and Dust Size Distributions of Predicted Virtual Showers}
\label{app:rad-act-size}

\slugcomment{See Figure~\ref{fig:a1} for a sample. All figures are separately attached to this arXiv submission.}

\clearpage

\clearpage

\clearpage

\clearpage

\clearpage

\clearpage

\clearpage

\renewcommand\thefigure{\thesection.\arabic{figure}}    
\setcounter{figure}{0}

\begin{deluxetable}{lcccccccccccccc}
\rotate
\tabletypesize{\scriptsize}
\tablecolumns{15}
\tablewidth{0pc}
\tablecaption{Objects that are capable to produce CMOR-detectable annual meteor activities. Listed are the properties of the parent (absolute magnitude $H$, Tisserand parameter with respect to Jupiter, $T_\mathrm{J}$, Minimum Orbit Intersection Distance (MOID) with respect to the Earth, orbital chaotic timescale $\tau_\mathrm{parent}$), dynamical properties of the hypothetical meteoroid stream (stream age $\tau_\mathrm{stream}$, encircling time $\tau_\mathrm{enc}$), and calculated meteor activities at ascending node \textohm and/or descending node \textmho (including the time of activity in solar longitude $\lambda_\odot$, radiant in J2000 sun-centered ecliptic coordinates, $\lambda-\lambda_\odot$ and $\beta$, radiant size $\sigma_\mathrm{rad}$, geocentric speed $v_\mathrm{g}$, and meteoroid flux $\mathcal{F}$ derived from median JFC model.}
\tablehead{
\colhead{} & \multicolumn{4}{c}{Parent} & \colhead{} & \multicolumn{9}{c}{Stream} \\
\cline{2-5} \cline{7-15} \\
\colhead{Parent} & \colhead{$H$} & \colhead{MOID} & \colhead{$T_\mathrm{J}$} & \colhead{$\tau_\mathrm{parent}$} & \colhead{} & \colhead{$\tau_\mathrm{stream}$} & \colhead{$\tau_\mathrm{enc}$} & \colhead{Node} & \colhead{$\lambda_\odot$} & \colhead{$\lambda-\lambda_\odot$} & \colhead{$\beta$} & \colhead{$\sigma_\mathrm{rad}$} & \colhead{$v_\mathrm{g}$} & \colhead{$\mathcal{F}_\mathrm{CMOR}$} \\
\colhead{} & \colhead{} & \colhead{(AU)} & \colhead{} & \colhead{(yr)} & \colhead{} & \colhead{(yr)} & \colhead{(yr)} & \colhead{} & \colhead{} & \colhead{} & \colhead{} & \colhead{} & \colhead{($\mathrm{km~s^{-1}}$)} & \colhead{($\mathrm{km^{-2}~hr^{-1}}$)} \\
}
\startdata
(3360) Syrinx & 15.9 & 0.108 & 2.965 & 5400 & & 4650 & $ 500 $ & \textmho & $ 212 ^\circ \pm 2 ^\circ $ & $ 357 ^\circ $ & $ + 23 ^\circ $ & $ \pm 1 ^\circ $ & $ 24.9 \pm 0.3 $ & 0.001 \\
(16960) 1998 QS$_{52}$ & 14.3 & 0.015 & 3.000 & $ >10000 $ & & 10000 & $ 100 $ & \textohm & $ 83 ^\circ \pm 1 ^\circ $ & $ 344 ^\circ $ & $ -13 ^\circ $ & $ \pm 1 ^\circ $ & $ 30.8 \pm 0.1 $ & 2.651 \\
(137427) 1999 TF$_{211}$ & 15.2 & 0.020 & 2.968 & 5300 & & $ 550 $ & $ 200 $ & \textmho & $ 348 ^\circ \pm 1 ^\circ $ & $ 345 ^\circ $ & $ + 81 ^\circ $ & $ \pm 1 ^\circ $ & $ 24.2 \pm 0.2 $ & 0.002 \\
(139359) 2001 ME$_1$ & 16.6 & 0.012 & 2.674 & 4900 & & $ 700 $ & $ 200 $ & \textmho & $ 92 ^\circ \pm 2 ^\circ $ & $ 191 ^\circ $ & $ + 4 ^\circ $ & $ \pm 1 ^\circ $ & $ 29.9 \pm 0.3 $ & 0.021 \\
(192642) 1999 RD$_{32}$ & 16.3 & 0.050 & 2.872 & 2500 & & $ 800 $ & $ 100 $ & \textohm & $ 155 ^\circ \pm 3 ^\circ $ & $ 2 ^\circ $ & $ -9 ^\circ $ & $ \pm 1 ^\circ $ & $ 22.8 \pm 0.6 $ & 0.001 \\
(196256) 2003 EH$_1$ & 16.2 & 0.212 & 2.065 & 1300 & & $ 300 $ & $ 200 $ & \textmho & $ 283 ^\circ \pm 1 ^\circ $ & $ 275 ^\circ $ & $ + 63 ^\circ $ & $ \pm 1 ^\circ $ & $ 41.6 \pm 0.2 $ & 0.012 \\
(247360) 2001 XU & 19.2 & 0.005 & 2.749 & $ >10000 $ & & 6700 & $ 100 $ & \textmho & $ 262 ^\circ \pm 1 ^\circ $ & $ 191 ^\circ $ & $ + 17 ^\circ $ & $ \pm 1 ^\circ $ & $ 29.1 \pm 0.2 $ & 0.773 \\
(248590) 2006 CS & 16.5 & 0.105 & 2.441 & 2000 & & $ 1800 $ & $ 200 $ & \textohm & $ 352 ^\circ \pm 1 ^\circ $ & $ 305 ^\circ $ & $ -77 ^\circ $ & $ \pm 1 ^\circ $ & $ 30.6 \pm 0.4 $ & 0.419 \\
(297274) 1996 SK & 16.8 & 0.004 & 2.968 & 2200 & & $ 1250 $ & $ 200 $ & \textmho & $ 204 ^\circ \pm 3 ^\circ $ & $ 184 ^\circ $ & $ + 2 ^\circ $ & $ \pm 1 ^\circ $ & $ 24.4 \pm 0.6 $ & 0.009 \\
(307005) 2001 XP$_1$ & 18.0 & 0.016 & 2.560 & $ >10000 $ & & $ 10000 $ & $ 200 $ & \textmho & $ 268 ^\circ \pm 1 ^\circ $ & $ 191 ^\circ $ & $ + 50 ^\circ $ & $ \pm 1 ^\circ $ & $ 28.5 \pm 0.0 $ & 0.781 \\
(399457) 2002 PD$_{43}$ & 19.1 & 0.029 & 2.439 & $ >10000 $ & & $ 300 $ & $ 300 $ & \textohm & $ 130 ^\circ \pm 3 ^\circ $ & $ 334 ^\circ $ & $ -8 ^\circ $ & $ \pm 1 ^\circ $ & $ 39.1 \pm 0.6 $ & 0.002 \\
(401857) 2000 PG$_3$ & 16.1 & 0.210 & 2.550 & 3200 & & $ 2150 $ & $ 100 $ & \textohm & $ 176 ^\circ \pm 2 ^\circ $ & $ 192 ^\circ $ & $ -12 ^\circ $ & $ \pm 1 ^\circ $ & $ 30.2 \pm 0.3 $ & 0.006 \\
(436329) 2010 GX$_{62}$ & 20.1 & 0.014 & 2.756 & $ >10000 $ & & $ 600 $ & $ 200 $ & \textohm & $ 25 ^\circ \pm 2 ^\circ $ & $ 19 ^\circ $ & $ -50 ^\circ $ & $ \pm 2 ^\circ $ & $ 18.7 \pm 0.3 $ & 0.003 \\
(442037) 2010 PR$_{66}$ & 19.3 & 0.002 & 2.818 & 2800 & & $ 1200 $ & $ 200 $ & \textohm & $ 114 ^\circ \pm 4 ^\circ $ & $ 25 ^\circ $ & $ -40 ^\circ $ & $ \pm 4 ^\circ $ & $ 16.4 \pm 0.2 $ & 0.001 \\
(451124) 2009 KC$_3$ & 18.0 & 0.006 & 2.728 & 4300 & & $ 900 $ & $ 700 $ & \textohm & $ 162 ^\circ \pm 2 ^\circ $ & $ 46 ^\circ $ & $ -31 ^\circ $ & $ \pm 2 ^\circ $ & $ 12.6 \pm 0.3 $ & 0.001 \\
1999 LT$_1$ & 17.6 & 0.095 & 2.586 & 2100 & & $ 1800 $ & $ 200 $ & \textmho & $ 67 ^\circ \pm 1 ^\circ $ & $ 343 ^\circ $ & $ + 78 ^\circ $ & $ \pm 1 ^\circ $ & $ 25.9 \pm 0.5 $ & 0.544 \\
2001 HA$_4$ & 17.7 & 0.018 & 2.772 & $ >10000 $ & & $ 4250 $ & $ 200 $ & \textohm & $ 179 ^\circ \pm 2 ^\circ $ & $ 184 ^\circ $ & $ -20 ^\circ $ & $ \pm 1 ^\circ $ & $ 25.0 \pm 0.2 $ & 0.122 \\
.. & .. & .. & .. & .. & & .. & .. & \textmho & $ 360 ^\circ \pm 2 ^\circ $ & $ 357 ^\circ $ & $ + 21 ^\circ $ & $ \pm 1 ^\circ $ & $ 24.8 \pm 0.3 $ & 0.028 \\
2002 EV$_{11}$ & 20.0 & 0.047 & 3.046 & $ >10000 $ & & $ 6600 $ & $ 150 $ & \textohm & $ 355 ^\circ \pm 3 ^\circ $ & $ 339 ^\circ $ & $ -6 ^\circ $ & $ \pm 1 ^\circ $ & $ 33.5 \pm 0.6 $ & 0.013 \\
2003 BK$_{47}$ & 17.8 & 0.026 & 2.857 & 8500 & & $ 4750 $ & $ 200 $ & \textmho & $ 133 ^\circ \pm 3 ^\circ $ & $ 169 ^\circ $ & $ + 36 ^\circ $ & $ \pm 2 $ & $ 19.5 \pm 0.2 $ & 0.006 \\
2003 CG$_{11}$ & 20.5 & 0.018 & 2.900 & $ >10000 $ & & $ 7200 $ & $ 200 $ & \textmho & $ 134 ^\circ \pm 2 ^\circ $ & $ 1 ^\circ $ & $ + 30 ^\circ $ & $ \pm 1 ^\circ $ & $ 22.9 \pm 0.2 $ & 0.057 \\
2003 OV & 18.3 & 0.082 & 2.987 & $ >10000 $ & & $ 6300 $ & $ 100 $ & \textohm & $ 108 ^\circ \pm 4 ^\circ $ & $ 194 ^\circ $ & $ -5 ^\circ $ & $ \pm 1 ^\circ $ & $ 30.0 \pm 0.8 $ & 0.015 \\
.. & .. & .. & .. & .. & & .. & .. & \textmho & $ 346 ^\circ \pm 4 ^\circ $ & $ 346 ^\circ $ & $ + 5 ^\circ $ & $ \pm 1 ^\circ $ & $ 29.5 \pm 0.9 $ & 0.008 \\
2004 BZ$_{74}$ & 18.1 & 0.032 & 2.369 & 7900 & & $ 3750 $ & $ 100 $ & \textohm & $ 60 ^\circ \pm 1 ^\circ $ & $ 192 ^\circ $ & $ -11 ^\circ $ & $ \pm 1 ^\circ $ & $ 32.0 \pm 0.2 $ & 0.044 \\
2004 CK$_{39}$ & 19.2 & 0.068 & 2.991 & $ >10000 $ & & $ 9350 $ & $ 150 $ & \textohm & $ 197 ^\circ \pm 3 ^\circ $ & $ 348 ^\circ $ & $ - 11 ^\circ $ & $ \pm 1 ^\circ $ & $ 29.1 \pm 0.4 $ & 0.002 \\
.. & .. & .. & .. & .. & & .. & .. & \textmho & $ 334 ^\circ \pm 2 ^\circ $ & $ 191 ^\circ $ & $ + 11 ^\circ $ & $ \pm 1 ^\circ $ & $ 29.1 \pm 0.4 $ & 0.010 \\
2004 TG$_{10}$ & 19.4 & 0.022 & 2.992 & 6600 & & $ 6100 $ & $ 400 $ & \textohm & $ 102 ^\circ \pm 2 ^\circ $ & $ 346 ^\circ $ & $ -3 ^\circ $ & $ \pm 1 ^\circ $ & $ 30.1 \pm 0.6 $ & 0.094 \\
.. & .. & .. & .. & .. & & .. & .. & \textmho & $ 223 ^\circ \pm 3 ^\circ $ & $ 194 ^\circ $ & $ + 3 ^\circ $ & $ \pm 1 ^\circ $ & $ 30.0 \pm 0.5 $ & 0.065 \\
2005 FH & 17.7 & 0.038 & 2.821 & 6200 & & $ 8100 $ & $ 150 $ & \textohm & $ 328 ^\circ \pm 2 ^\circ $ & $ 3 ^\circ $ & $ -58 ^\circ $ & $ \pm 2 ^\circ $ & $ 22.5 \pm 0.2 $ & 0.012 \\
2005 UN$_{157}$ & 18.2 & 0.420 & 2.581 & 6000 & & $ 3750 $ & $ 200 $ & \textohm & $ 175 ^\circ \pm 2 ^\circ $ & $ 339 ^\circ $ & $ -13 ^\circ $ & $ \pm 1 ^\circ $ & $ 36.4 \pm 0.4 $ & 0.008 \\
.. & .. & .. & .. & .. & & .. & .. & \textmho & $ 261 ^\circ \pm 3 ^\circ $ & $ 203 ^\circ $ & $ + 13 ^\circ $ & $ \pm 1 ^\circ $ & $ 36.9 \pm 0.5 $ & 0.003 \\
2005 WY$_{55}$ & 20.7 & 0.004 & 3.042 & 4700 & & $ 2500 $ & $ 1200 $ & \textohm & $ 70 ^\circ \pm 3 ^\circ $ & $ 7 ^\circ $ & $ -12 ^\circ $ & $ \pm 1 ^\circ $ & $ 19.6 \pm 0.6 $ & 0.012 \\
2006 AL$_8$ & 18.4 & 0.056 & 2.159 & 2900 & & $ 7300 $ & $ 100 $ & \textmho & $ 312 ^\circ \pm 2 ^\circ $ & $ 339 ^\circ $ & $ + 24 ^\circ $ & $ \pm 1 ^\circ $ & $ 36.0 \pm 0.4 $ & 0.013 \\
2006 KK$_{21}$ & 20.4 & 0.033 & 2.605 & 5000 & & $ 4000 $ & $ 100 $ & \textohm & $ 51 ^\circ \pm 2 ^\circ $ & $ 347 ^\circ $ & $ -11 ^\circ $ & $ \pm 1 ^\circ $ & $ 30.5 \pm 0.4 $ & 0.022 \\
.. & .. & .. & .. & .. & & .. & .. & \textmho & $ 180 ^\circ \pm 2 ^\circ $ & $ 192 ^\circ $ & $ + 10 ^\circ $ & $ \pm 1 ^\circ $ & $ 30.2 \pm 0.3 $ & 0.031 \\
2007 CA$_{19}$ & 17.6 & 0.019 & 2.679 & 4300 & & $ 300 $ & $ 100 $ & \textohm & $ 354 ^\circ \pm 1 ^\circ $ & $ 185 ^\circ $ & $ -10 ^\circ $ & $ \pm 1 ^\circ $ & $ 27.1 \pm 0.2 $ & 0.018 \\
.. & .. & .. & .. & .. & & .. & .. & \textmho & $ 190 ^\circ \pm 1 ^\circ $ & $ 355 ^\circ $ & $ + 10 ^\circ $ & $ \pm 1 ^\circ $ & $ 27.0 \pm 0.1 $ & 0.010 \\
2008 SV$_{11}$ & 18.4 & 0.018 & 2.957 & 8400 & & $ 3300 $ & $ 200 $ & \textmho & $ 8 ^\circ \pm 4 ^\circ $ & $ 10 ^\circ $ & $ + 15 ^\circ $ & $ \pm 3 $ & $ 18.6 \pm 0.5 $ & 0.005 \\
2008 YZ$_{28}$ & 20.0 & 0.094 & 2.969 & 3300 & & $ 4000 $ & $ 200 $ & \textmho & $ 270 ^\circ \pm 6 ^\circ $ & $ 358 ^\circ $ & $ + 54 ^\circ $ & $ \pm 5 $ & $ 23.8 \pm 0.4 $ & 0.001 \\
2009 HD$_{21}$ & 18.2 & 0.015 & 2.881 & 6900 & & $ 1400 $ & $ 100 $ & \textmho & $ 180 ^\circ \pm 3 ^\circ $ & $ 20 ^\circ $ & $ + 41 ^\circ $ & $ \pm 3 $ & $ 17.7 \pm 0.2 $ & 0.005 \\
2009 SG$_{18}$ & 17.8 & 0.025 & 2.313 & 8500 & & $ 9250 $ & $ 200 $ & \textmho & $ 177 ^\circ \pm 1 ^\circ $ & $ 237 ^\circ $ & $ + 70 ^\circ $ & $ \pm 1 ^\circ $ & $ 34.1 \pm 0.3 $ & 0.172 \\
2009 WN$_{25}$ & 18.4 & 0.114 & 1.959 & 2700 & & $ 100 $ & $ 400 $ & \textmho & $ 232 ^\circ \pm 1 ^\circ $ & $ 271 ^\circ $ & $ + 63 ^\circ $ & $ \pm 1 ^\circ $ & $ 41.7 \pm 0.1 $ & 1.034 \\
2010 JL$_{33}$ & 17.7 & 0.033 & 2.910 & 4100 & & $ 4000 $ & $ 100 $ & \textohm & $ 250 ^\circ \pm 5 ^\circ $ & $ 10 ^\circ $ & $ -8 ^\circ $ & $ \pm 2 ^\circ $ & $ 19.0 \pm 1.1 $ & 0.001 \\
2010 XC$_{11}$ & 18.7 & 0.030 & 2.792 & 6400 & & $ 850 $ & $ 700 $ & \textohm & $ 282 ^\circ \pm 2 ^\circ $ & $ 192 ^\circ $ & $ -7 ^\circ $ & $ \pm 1 ^\circ $ & $ 29.9 \pm 0.4 $ & 0.002 \\
2011 GH$_3$ & 18.5 & 0.149 & 3.020 & 8600 & & $ 6850 $ & $ 200 $ & \textohm & $ 237 ^\circ \pm 1 ^\circ $ & $ 357 ^\circ $ & $ -9 ^\circ $ & $ \pm 1 ^\circ $ & $ 24.6 \pm 0.2 $ & 0.011 \\
.. & .. & .. & .. & .. & & .. & .. & \textmho & $ 49 ^\circ \pm 1 ^\circ $ & $ 183 ^\circ $ & $ + 9 ^\circ $ & $ \pm 1 ^\circ $ & $ 24.7 \pm 0.2 $ & 0.013 \\
2011 GN$_{44}$ & 18.3 & 0.009 & 2.922 & $ >10000 $ & & $ 10000 $ & $ 200 $ & \textohm & $ 196 ^\circ \pm 1 ^\circ $ & $ 318 ^\circ $ & $ -65 ^\circ $ & $ \pm 1 ^\circ $ & $ 32.6 \pm 0.1 $ & 5.829 \\
2012 BU$_{61}$ & 21.3 & 0.027 & 2.933 & 8400 & & $ 1700 $ & $ 1100 $ & \textohm & $ 101 ^\circ \pm 4 ^\circ $ & $ 180 ^\circ $ & $ -6 ^\circ $ & $ \pm 1 ^\circ $ & $ 23.1 \pm 0.9 $ & 0.002 \\
.. & .. & .. & .. & .. & & .. & .. & \textmho & $ 280 ^\circ \pm 4 ^\circ $ & $ 359 ^\circ $ & $ + 6 ^\circ $ & $ \pm 1 ^\circ $ & $ 23.5 \pm 1.0 $ & 0.001 \\
2012 FZ$_{23}$ & 18.2 & 0.020 & 2.367 & 4400 & & $ 1250 $ & $ 200 $ & \textohm & $ 359 ^\circ \pm 1 ^\circ $ & $ 269 ^\circ $ & $ -61 ^\circ $ & $ \pm 1 ^\circ $ & $ 41.7 \pm 0.2 $ & 0.369 \\
2012 HG$_8$ & 19.7 & 0.004 & 2.967 & 4000 & & $ 2800 $ & $ 100 $ & \textmho & $ 215 ^\circ \pm 4 ^\circ $ & $ 182 ^\circ $ & $ + 36 ^\circ $ & $ \pm 3 $ & $ 23.7 \pm 0.3 $ & 0.013 \\
2012 TO$_{139}$ & 19.7 & 0.001 & 2.759 & 4800 & & $ 300 $ & $ 100 $ & \textohm & $ 290 ^\circ \pm 4 ^\circ $ & $ 196 ^\circ $ & $ -4 ^\circ $ & $ \pm 1 ^\circ $ & $ 33.1 \pm 0.8 $ & 0.001 \\
.. & .. & .. & .. & .. & & .. & .. & \textmho & $ 179 ^\circ \pm 1 ^\circ $ & $ 345 ^\circ $ & $ + 3 ^\circ $ & $ \pm 1 ^\circ $ & $ 32.6 \pm 0.1 $ & 0.196 \\
2015 TB$_{145}$ & 19.9 & 0.002 & 2.964 & $ >10000 $ & & $ 10000 $ & $ 100 $ & \textohm & $ 217 ^\circ \pm 1 ^\circ $ & $ 204 ^\circ $ & $ -24 ^\circ $ & $ \pm 1 ^\circ $ & $ 34.9 \pm 0.2 $ & 1.738 \\
\enddata
\label{tbl:can-met-ev}
\end{deluxetable}

\clearpage

\begin{deluxetable}{lcccccccccccc}
\rotate
\tabletypesize{\scriptsize}
\tablecolumns{13}
\tablewidth{0pc}
\tablecaption{Orbits and radiant characteristics of the possible meteor activity associated with 2009 WN$_{25}$, 2011 BE$_{38}$ and 2012 BU$_{61}$. Listed are perihelion distance $q$, eccentricity $e$, inclination $i$, longitude of ascending node $\Omega$ and argument of perihelion $\omega$ for the parent (taken from JPL~31, 28 and 15 for the respective parent) and the meteor shower from the given reference. The uncertainties in the orbital elements for the parents are typically in the order of $10^{-5}$ to $10^{-8}$ in their respective units and are not shown. Epochs are in J2000. Shwon are the absolute magnitude of the parent as well as the expected number of NEOs with $H<18$ and $H<22$ which are expected to have $D'<D'_0$ relative to that of the proposed parent. Values of $\langle X \rangle$ near or larger than 1 suggest that the association is not statistically significant.}
\tablehead{
\colhead{} & \multicolumn{5}{c}{Orbital elements (J2000)} & \colhead{} & \multicolumn{3}{c}{Geocentric radiant (J2000)} & \colhead{} & \multicolumn{2}{c}{$\langle X \rangle$} \\
\cline{2-6} \cline{8-10} \cline{12-13} \\
\colhead{} & \colhead{$q$} & \colhead{$e$} & \colhead{$i$} & \colhead{$\Omega$} & \colhead{$\omega$} & \colhead{} & \colhead{$\lambda-\lambda_\odot$} & \colhead{$\beta$} & \colhead{$v_\mathrm{g}$} & \colhead{} & \colhead{$H<18$} & \colhead{$H<22$} \\
\colhead{} & \colhead{(AU)} & \colhead{} & \colhead{} & \colhead{} & \colhead{} & \colhead{} & \colhead{} & \colhead{} & \colhead{($\mathrm{km~s^{-1}}$)} & \colhead{} & \colhead{} \\
}
\startdata
\sidehead{2009 WN$_{25}$ ($H=18.4$) -- November i Draconids}
2009 WN$_{25}$ & 1.10238 & 0.66278 & $71.986^\circ$ & $232.086^\circ$ & $180.910^\circ$ & & $271^\circ$ & $+63^\circ$ & $41.7$ & & & \\
 & & & & & & & $\pm1^\circ$ & $\pm1^\circ$ & $\pm0.1$ & & & \\
Shower prediction -- this work & $0.987$ & $0.619$ & $73.6^\circ$ & $238.0^\circ$ & $184.6^\circ$ & & $267.6^\circ$ & $+62.0^\circ$ & $41.4$ & & 0.001--0.05 & 0.02--0.6 \\
 & $\pm0.002$ & $\pm 0.133$ & $7\pm 2.3^\circ$ & $\pm 0.5^\circ$ & $\pm 2.9^\circ$ & & $\pm 0.1^\circ$ & $\pm 0.1^\circ$ & $\pm 0.1$ & & & \\
Observed shower -- \citet{Brown2010} & 0.9874 & 0.737 & $74.9^\circ$ & $241.0^\circ$ & $181.09^\circ$ & & $270.1^\circ$ & $+62.5^\circ$ & 43 & & 0.003 & 0.04 \\
Observed shower -- \citet{Jenniskens2016} & 0.973 & 0.734 & $72.9^\circ$ & $254.4^\circ$ & $194.7^\circ$ & & $260.9^\circ$ & $+63.2^\circ$ & 41.9 & & 0.4 & 6 \\
\hline
\sidehead{2002 AU$_5$ ($H=17.8$) \& 2012 BU$_{61}$ ($H=21.5$) -- Daytime $\xi$ Sagittariids (XSA) \& Daytime Scutids (JSC)\tablenotemark{b}}
2002 AU$_5$ & 0.40301 & 0.75531 & $9.256^\circ$ & $354.989^\circ$ & $21.261^\circ$ & & $359^\circ$ & $+6^\circ$ & $23.5$ & & & \\
 & & & & & & & $\pm1^\circ$ & $\pm1^\circ$ & $\pm1.0$ & & & \\
2012 BU$_{61}$ & 0.55333 & 0.78023 & $5.277^\circ$ & $297.700^\circ$ & $72.461^\circ$ & & $359^\circ$ & $+6^\circ$ & $23.5$ & & & \\
 & & & & & & & $\pm1^\circ$ & $\pm1^\circ$ & $\pm1.0$ & & & \\
XSA prediction -- this work & $0.46$ & $0.77$ & $5.9^\circ$ & $291.0^\circ$ & $76.6^\circ$ & & $352.6^\circ$ & $+6.4^\circ $ & 25.2 & & AU$_5$: 0.8--1.2 & AU$_5$: 10--15 \\
 & $\pm0.02$ & $\pm 0.04$ & $\pm 1.0^\circ$ & $\pm 0.5^\circ$ & $\pm 2.1^\circ$ & & $\pm 0.1^\circ$ & $\pm 0.1^\circ$ & $\pm0.1$ & & BU$_{61}$: 0.05--0.2 & BU$_{61}$: 0.6--2 \\
XSA observation -- \citet{Sekanina1976g} & $0.29$ & $0.74$ & $1.1^\circ$ & $304.9^\circ$ & $46.9^\circ$ & & $338.0^\circ$ & $+0.9^\circ$ & 24.4 & & AU$_5$: 12--13 & AU$_5$: 156--164 \\
 & $\pm0.01$ & $\pm 0.02$ & $\pm 0.7^\circ$ & $\pm 1.4^\circ$ & $\pm 1.8^\circ$ & & $\pm 1.0^\circ$ & $\pm 0.6^\circ$ & & & BU$_{61}$: 8--10 & BU$_{61}$: 101--126 \\
XSA observation -- \citet{Brown2010} & 0.4708 & 0.784 & $6.0^\circ$ & $288.0^\circ$ & $79.31^\circ$ & & $353.9^\circ$ & $+6.6^\circ$ & 25.3 & & AU$_5$: 1.0 & AU$_5$: 13 \\
 & & & & & & & & & & & BU$_{61}$: 0.04 & BU$_{61}$: 0.6 \\
JSC observation -- \citet{Sekanina1973b} & $0.55$ & $0.77$ & $12.4^\circ$ & $280.4^\circ$ & $89.4^\circ$ & & $358.5^\circ$ & $+15.4^\circ$ & 24.1 & & AU$_5$: 1.8--2.3 & AU$_5$: 23--29 \\
 & $\pm0.01$ & $\pm 0.02$ & $\pm 1.4^\circ$ & $\pm 0.5^\circ$ & $\pm 1.1^\circ$ & & $\pm 0.4^\circ$ & $\pm 1.8^\circ$ & & & BU$_{61}$: 0.05--0.3 & BU$_{61}$: 0.3--4 \\
\enddata
\tablenotetext{a}{Ascending node.}
\tablenotetext{b}{Descending node.}
\label{tbl:met-ev-positive}
\end{deluxetable}

\clearpage

\begin{deluxetable}{lccccc}
\rotate
\tablecolumns{6}
\tablewidth{0pc}
\tablecaption{Previously proposed associations that are not reproduced in this work. Only objects that are in our initial 407-object list are included. ``Established showers'' means confirmed meteor showers in the IAU catalog, not established parent-shower linkages (likewise for unestablished showers). Listed are the absolute magnitude of the parent $H$, sources where the linkage was proposed, orbital elements, and $\langle X \rangle$ for the NEO population of $H<18$ and $H<22$.}
\tablehead{
\colhead{Shower} & \colhead{Proposed parent} & \colhead{$H$} & \colhead{Reference} & \colhead{$\langle X \rangle_{H<18}$} & \colhead{$\langle X \rangle_{H<22}$} \\
}
\startdata
\sidehead{Established showers:}
Corvids & (14827) Hypnos & 18.3 & O87, J16, JPL~49 & 1.4 & 18 \\
.. & (374038) 2004 HW & 17.0 & Je06, J16, JPL~60 & 0.1 & 1.4 \\
Daytime April Piscids & 2003 MT$_9$ & 18.6 & B09, B10, JPL~37 & 0.9 & 11 \\
.. & (401857) 2000 PG$_3$\tablenotemark{a} & 16.1 & B09, B10, JPL~43 & 34 & 432 \\
.. & 2002 JC$_9$\tablenotemark{a} & 18.5 & B09, B10, JPL~26 & 10 & 121 \\
$\kappa$ Cygnids & (153311) 2001 MG$_1$ & 17.2 & Jo06, J16, JPL~63 & 0.8 & 10 \\
.. & (361861) 2008 ED$_{69}$ & 17.0 & J08, J16, JPL~36 & 1.7 & 21 \\
Northern $\iota$ Aquariids & 2003 MT$_9$ & 18.6 & B09, J16, JPL~37 & 9 & 114 \\
$\psi$ Cassiopeiids & (5496) 1973 NA & 16.0 & P92, J16, JPL~51 & 0.2 & 1.9 \\
\hline
\sidehead{Unestablished showers:}
66 Draconids & 2001 XQ & 19.2 & S14a, JPL~14 & 0.003 & 0.04 \\
August $\theta$ Aquillids & 2004 MB$_6$ & 19.5 & K14, K15, JPL~17 & 16 & 203 \\
Daytime April Cetids & 2003 MT$_9$ & 18.6 & K67, B09, JPL~37 & 0.6 & 8 \\
Daytime c Aquariids & (206910) 2004 NL$_8$ & 17.1 & G75, Je06, JPL~108 & 7 & 89 \\
Daytime $\delta$ Scorpiids & 2003 HP$_{32}$\tablenotemark{b} & 19.6 & N64, B15, JPL~17 & 0.6 & 8 \\
.. & 2007 WY$_3$\tablenotemark{b} & 18.2 & N64, B15, JPL~18 & 6 & 75 \\
$\delta$ Mensids & (248590) 2006 CS & 16.5 & Je06, JPL~48 & 0.001 & 0.02 \\
$\eta$ Virginids & 2007 CA$_{19}$ & 17.6 & B15, J16, JPL~56 & 3.3 & 42 \\
$\gamma$ Piscids & 6344 P-L & 20.4 & T89, Je06, JPL~16 & 51 & 648 \\
$\gamma$ Triangulids & 2002 GZ$_8$ & 18.2 & P94, Je06, JPL~33 & 3.1 & 39 \\
$\iota$ Cygnids & 2001 SS$_{287}$ & 18.3 & A13, JPL~21 & 0.1 & 1.2 \\
$\kappa$ Cepheids & 2009 SG$_{18}$ & 17.8 & S15, JPL~22 & 0.0004 & 0.006 \\
$\lambda$ Cygnids & (189263) 2005 CA & 15.6 & T89, Je06, JPL~51 & 15 & 185 \\
Northern $\delta$ Leonids & (192642) 1999 RD$_{32}$ & 16.3 & L71, Je06, JPL~125 & 0.7 & 9 \\
Northern $\delta$ Piscids & (401857) 2000 PG$_3$\tablenotemark{a} & 16.1 & B09, J16, JPL~43 & 24 & 302 \\
.. & 2002 JC$_9$\tablenotemark{a} & 18.5 & B09, J16, JPL~26 & 8 & 107 \\
Northern $\gamma$ Virginids & 2002 FC & 18.9 & T89, Je06, JPL~52 & 0.2 & 2.2 \\
Northern $\sigma$ Sagittariids\tablenotemark{c} & (139359) 2001 ME$_1$ & 16.6 & S76, Je06, JPL~71 & 2.3 & 29 \\
Southern $\alpha$ Leonids & (172678) 2003 YM$_{137}$ & 18.7 & Je06, JPL~51 & 13 & 166 \\
Southern $\delta$ Piscids & (401857) 2000 PG$_3$\tablenotemark{a} & 16.1 & B09, J16, JPL~43 & 15 & 195 \\
.. & 2002 JC$_9$\tablenotemark{a} & 18.5 & B09, J16, JPL~26 & 20 & 251 \\
Southern $\iota$ Aquariids & 2003 MT$_9$ & 18.6 & B08, B09, JPL~37 & 0.6 & 7 \\
$\zeta^1$ Cancrids & 2012 TO$_{139}$ & 19.7 & S14b, JPL~13 & 0.08 & 1.0 \\
\enddata
\tablecomments{Abbreviation of references: A13 -- \citet{Andreic2013a}; B08 -- \citet{Brown2008}; B09 -- \citet{Babadzhanov2009b}; B10 -- \citet{Brown2010}; B15 -- \citet{Babadzhanov2015c}; G75 -- \citet{Gartrell1975b}; Je06 -- \citet{Jenniskens2006b}; Jo06 -- \citet{Jones2006id}; J08 -- \citet{Jenniskens2008}; J16 -- \citet{Jenniskens2016}; K14 -- \citet{Kornovs2014b}; K15 -- \citet{Kokhirova2015c}; K67 -- \citet{Kashcheyev1967a}; L71 -- \citet{Lindblad1971a}; N64 -- \citet{Nilsson1964c}; O87 -- \citet{Olsson-Steel1987}; P92 -- \citet{Porubcan1992e}; \citet{Porubcan1994}; S14a -- \citet{vSegon2014d}; S14b -- \citet{vSegon2014g}; S76 -- \citet{Sekanina1976g}; T89 -- \citet{Terentjeva1989a}}
\tablenotetext{a}{Objects/streams thought be belonged to the same complex.}
\tablenotetext{b}{Objects/streams thought be belonged to the same complex.}
\tablenotetext{c}{Called $\sigma$ Carpricornids in \citet{Sekanina1976g}.}
\label{tbl:negative}
\end{deluxetable}

\clearpage

\begin{deluxetable}{lcccccccc}
\rotate
\tablecolumns{9}
\tablewidth{0pc}
\tablecaption{Predicted meteor outbursts from virtual young meteoroid trails from the shower parents. Shown are the age of the trail, period of expected activity (in date and solar longitude, $\lambda_\odot$, rounded to the nearest $1^\circ$ solar longitude), radiant (in J2000 sun-centered ecliptic coordinates, $\lambda-\lambda_\odot$ and $\beta$), geocentric speed ($v_\mathrm{g}$), and estimated meteoroid flux derived from median JFC model.}
\tablehead{
\colhead{Parent} & \colhead{$\tau_\mathrm{enc}$} & \colhead{Date} & \colhead{Ejection} & \colhead{$\lambda_\odot$} & \colhead{$\lambda-\lambda_\odot$} & \colhead{$\beta$} & \colhead{$v_\mathrm{g}$} & \colhead{$\mathcal{F}_\mathrm{CMOR}$} \\
\colhead{} & \colhead{(yr)} & \colhead{(UT)} & \colhead{} & \colhead{} & \colhead{} & \colhead{} & \colhead{($\mathrm{km \cdot s^{-1}}$)} & \colhead{($\mathrm{hr^{-1} \cdot km^{-2}}$)}
}
\startdata
(139359) 2001 ME$_1$ & 200 & 2006 Jun. 24 & 1924--1967 & $ 93 ^\circ $ & $ 191 ^\circ $ & $ +4 ^\circ $ & 30.0 & 0.01 \\
(247360) 2001 XU & 100 & 2014 Dec. 14 & 1903--1993 & $ 263 ^\circ $ & $ 190 ^\circ $ & $ +17 ^\circ $ & 29.1 & 0.02 \\
(297274) 1996 SK & 200 & 2007 Apr. 17 & 1870--1903 & $ 27 ^\circ $ & $ 2 ^\circ $ & $ -3 ^\circ $ & 21.5 & 0.01 \\
(435159) 2007 LQ$_{19}$ & 200 & 2002 Jul. 13 & 1801--1978 & $ 111 ^\circ $ & $ 139 ^\circ $ & $ +50 ^\circ $ & 14.1 & 0.01 \\
.. & .. & 2006 Jul. 13 & 1801--2003 & $ 111 ^\circ $ & $ 139 ^\circ $ & $ +50 ^\circ $ & 14.1 & 0.03 \\
.. & .. & 2007 Jul. 13 & 1805--2003 & $ 111 ^\circ $ & $ 139 ^\circ $ & $ +50 ^\circ $ & 14.1 & 0.04 \\
2001 HA$_4$ & 200 & 2005 Sep. 21 & 1807--1974 & $ 179 ^\circ $ & $ 184 ^\circ $ & $ -20 ^\circ $ & 24.5 & 2.71 \\
2005 WY$_{55}$ & 1200 & 2002 May 31\tablenotemark{\dag} & 994--1765 & $ 70 ^\circ $ & $ 8 ^\circ $ & $ -12 ^\circ $ & 19.3 & 0.16 \\
.. & .. & 2006 May 31\tablenotemark{\ddag} & 990--1753 & $ 70 ^\circ $ & $ 7 ^\circ $ & $ -12 ^\circ $ & 19.5 & 0.14 \\
.. & .. & 2010 May 31 & 875--1761 & $ 70 ^\circ $ & $ 8 ^\circ $ & $ -12 ^\circ $ & 19.4 & 0.06 \\
.. & .. & 2014 May 31 & 951--1725 & $ 70 ^\circ $ & $ 7 ^\circ $ & $ -12 ^\circ $ & 19.5 & 0.05 \\
2007 CA$_{19}$ & 100 & 2012 Mar. 14 & 1965--1993 & $ 354 ^\circ $ & $ 185 ^\circ $ & $ -10 ^\circ $ & 27.1 & 0.42 \\
2009 SG$_{18}$ & 200 & 2006 Sep. 20 & 1831--1852 & $ 178 ^\circ $ & $ 238 ^\circ $ & $ +70 ^\circ $ & 34.1 & 0.01 \\
.. & .. & 2015 Sep. 21 & 1920--1931 & $ 178 ^\circ $ & $ 239 ^\circ $ & $ +70 ^\circ $ & 34.3 & 0.01 \\
2012 BU$_{61}$ & 1100 & 2007 Jun. 29 & 1500--1788 & $ 97 ^\circ $ & $ 181 ^\circ $ & $ -6 ^\circ $ & 23.9 & 0.05 \\
2012 TO$_{139}$ & 100 & 2012 Sep. 21 & 1954--2008 & $ 179 ^\circ $ & $ 345 ^\circ $ & $ +4 ^\circ $ & 32.6 & 16.62 \\
2015 TB$_{145}$ & 100 & 2003 Oct. 31 & 1902--1997 & $ 218 ^\circ $ & $ 204 ^\circ $ & $ -24 ^\circ $ & 34.9 & 12.62 \\
.. & .. & 2004 Oct. 31 & 1908--1979 & $ 218 ^\circ $ & $ 204 ^\circ $ & $ -24 ^\circ $ & 35.0 & 0.01 \\
.. & .. & 2006 Oct. 31 & 1902--2000 & $ 218 ^\circ $ & $ 204 ^\circ $ & $ -24 ^\circ $ & 34.9 & 23.54 \\
.. & .. & 2009 Oct. 31 & 1902--2006 & $ 218 ^\circ $ & $ 204 ^\circ $ & $ -24 ^\circ $ & 34.9 & 42.86 \\
.. & .. & 2010 Oct. 31\tablenotemark{\ddag} & 1905--1991 & $ 218 ^\circ $ & $ 204 ^\circ $ & $ -24 ^\circ $ & 34.9 & 0.14 \\
.. & .. & 2012 Oct. 31 & 1930--2009 & $ 218 ^\circ $ & $ 204 ^\circ $ & $ -24 ^\circ $ & 34.8 & 10.95 \\
.. & .. & 2013 Oct. 31 & 1902--1982 & $ 218 ^\circ $ & $ 204 ^\circ $ & $ -24 ^\circ $ & 34.9 & 0.10 \\
.. & .. & 2014 Oct. 31 & 1911--1960 & $ 218 ^\circ $ & $ 204 ^\circ $ & $ -24 ^\circ $ & 35.0 & 0.01 \\
.. & .. & 2015 Oct. 31 & 1905--2015 & $ 218 ^\circ $ & $ 204 ^\circ $ & $ -24 ^\circ $ & 34.9 & 9.42 \\
\enddata
\tablenotetext{\dag}{CMOR not operational during the day.}
\tablenotetext{\ddag}{CMOR partially operational during the day.}
\label{tbl:can-met-ob}
\end{deluxetable}

\clearpage

\begin{deluxetable}{lcccccccccccc}
\rotate
\tablecolumns{13}
\tablewidth{0pc}
\tablecaption{Orbits and radiant characteristics of the possible meteor activity associated to (139359) 2001 ME$_1$. Listed are perihelion distance $q$, eccentricity $e$, inclination $i$, longitude of ascending node $\Omega$ and argument of perihelion $\omega$ for the parent (taken from JPL~71) and the meteor outburst in 2006 (derived from the corresponding wavelet maximum). The uncertainties in the orbital elements for the parents are typically in the order of $10^{-5}$ to $10^{-8}$ in their respective units and are not shown. Epochs are in J2000.}
\tablehead{
\colhead{} & \multicolumn{5}{c}{Orbital elements (J2000)} & \colhead{} & \multicolumn{3}{c}{Geocentric radiant (J2000)} & \colhead{} & \multicolumn{2}{c}{$\langle X \rangle$} \\
\cline{2-6} \cline{8-10} \cline{12-13} \\
\colhead{} & \colhead{$q$} & \colhead{$e$} & \colhead{$i$} & \colhead{$\Omega$} & \colhead{$\omega$} & \colhead{} & \colhead{$\lambda-\lambda_\odot$} & \colhead{$\beta$} & \colhead{$v_\mathrm{g}$} & \colhead{} & \colhead{$H<18$} & \colhead{$H<22$} \\
\colhead{} & \colhead{(AU)} & \colhead{} & \colhead{} & \colhead{} & \colhead{} & \colhead{} & \colhead{} & \colhead{} & \colhead{($\mathrm{km~s^{-1}}$)} & \colhead{} & \colhead{} \\
}
\startdata
(139359) 2001 ME$_1$ & 0.35512 & 0.86598 & $5.796^\circ$ & $86.506^\circ$ & $300.254^\circ$ & & $191^\circ$ & $+4^\circ$ & $30.0$ & & & \\
 & & & & & & & $\pm1^\circ$ & $\pm1^\circ$ & $\pm0.1$ & & & \\
2006 outburst & $0.32$ & $0.87$ & $4.7^\circ$ & $93.0^\circ$ & $298.8^\circ$ & & $193.0^\circ$ & $+3.5^\circ$ & $30.5$ & & 0.01--0.02 & 0.1--0.3 \\
 & $\pm0.02$ & $\pm0.03$ & $\pm1.4^\circ$ & $\pm0.5^\circ$ & $\pm2.2^\circ$ & & $\pm0.5^\circ$ & $\pm0.5^\circ$ & $\pm0.5$ & & & \\
\enddata
\label{tbl:cmor-139359-2006}
\end{deluxetable}

\end{CJK*}
\end{document}